\documentclass[aps,prd,floatfix,footinbib]{revtex4}
\usepackage{amsmath}
\usepackage{natbib}
\usepackage{epsfig}
\usepackage[mathscr]{eucal}
\usepackage{bbm}
\usepackage{bm}
\usepackage{hyperref}
\usepackage[toc]{appendix}

\newcommand{\half}{\frac{1}{2}}

\def\bra#1{\mathinner{\langle{#1}|}} 
\def\ket#1{\mathinner{|{#1}\rangle}}

\makeatletter

\newbox\swb@xone
\newbox\swb@xtwo
\newbox\swb@xthree
\newbox\swb@xfour
\newdimen\swdimen@ne
\newdimen\swdimentw@

\newcommand{\acontraction}[5][1ex]{%
  \mathchoice
    {\acontraction@\displaystyle{#2}{#3}{#4}{#5}{#1}}%
    {\acontraction@\textstyle{#2}{#3}{#4}{#5}{#1}}%
    {\acontraction@\scriptstyle{#2}{#3}{#4}{#5}{#1}}%
    {\acontraction@\scriptscriptstyle{#2}{#3}{#4}{#5}{#1}}}%
\newcommand{\acontraction@}[6]{%
  \setbox\swb@xone=\hbox{${}#1{}#2{}$}%
  \setbox\swb@xtwo=\hbox{${}#1{}#3{}$}%
  \setbox\swb@xthree=\hbox{${}#1{}#4{}$}%
  \setbox\swb@xfour=\hbox{${}#1{}#5{}$}%
  \swdimen@ne=\wd\swb@xtwo%
  \advance\swdimen@ne by \wd\swb@xfour%
  \divide\swdimen@ne by 2%
  \advance\swdimen@ne by \wd\swb@xthree%
  \vbox{%
    \hbox to 0pt{%
      \kern \wd\swb@xone%
      \kern 0.5\wd\swb@xtwo%
      \acontraction@@{\swdimen@ne}{#6}%
      \hss}%
    \vskip 0.5ex
    \vskip\ht\swb@xtwo}}

\newcommand{\acontraction@@}[3][0.05em]{%
  \hbox{%
    \vrule width #1 height 0pt depth #3%
    \vrule width #2 height 0pt depth #1%
    \vrule width #1 height 0pt depth #3%
    \relax}}
\let\contraction\acontraction

\newcommand{\bcontraction}[5][1ex]{%
  \mathchoice
    {\bcontraction@\displaystyle{#2}{#3}{#4}{#5}{#1}}%
    {\bcontraction@\textstyle{#2}{#3}{#4}{#5}{#1}}%
    {\bcontraction@\scriptstyle{#2}{#3}{#4}{#5}{#1}}%
    {\bcontraction@\scriptscriptstyle{#2}{#3}{#4}{#5}{#1}}}%
\newcommand{\bcontraction@}[6]{%
  \setbox\swb@xone=\hbox{${}#1{}#2{}$}%
  \setbox\swb@xtwo=\hbox{${}#1{}#3{}$}%
  \setbox\swb@xthree=\hbox{${}#1{}#4{}$}%
  \setbox\swb@xfour=\hbox{${}#1{}#5{}$}%
  \swdimen@ne=\wd\swb@xtwo%
  \advance\swdimen@ne by \wd\swb@xfour%
  \divide\swdimen@ne by 2%
  \advance\swdimen@ne by \wd\swb@xthree%
  \lower 0.5ex \vbox{%
    \hbox to 0pt{%
      \kern \wd\swb@xone%
      \kern 0.5\wd\swb@xtwo%
      \bcontraction@@{\swdimen@ne}{#6}%
      \hss}%
    }}

\newcommand{\bcontraction@@}[3][0.05em]{%
  \hbox{%
    \swdimentw@=#3
    \advance\swdimentw@ by -#1
    \vrule width #1 height 0pt depth #3%
    \lower\swdimentw@\hbox{\vrule width #2 height 0pt depth #1}%
    \vrule width #1 height 0pt depth #3%
    \relax}}

\makeatother

\begin{document}
\title{Generation of Circular Polarization of the Cosmic Microwave Background}

\author{Stephon Alexander} 
\affiliation{Department of Physics and Astronomy, Haverford College, Haverford, PA 19041, USA}
\author{Joseph Ochoa}
\affiliation{Department of Physics, Institute for Gravitation and the Cosmos, The Pennsylvania State University, 104 Davey Lab, University Park, PA 16802, USA}
\author{Arthur Kosowsky}
\affiliation{Department of Physics and Astronomy, University of Pittsburgh, Pittsburgh, PA 15208, USA}

\begin{abstract}
The standard cosmological model, which includes only Compton scattering photon interactions
at energy scales near recombination, results in zero primordial circular polarization of the 
cosmic microwave background. In this paper we consider a particular renormalizable and
gauge-invariant standard model extension coupling photons to an
external vector field via a Chern-Simons term, which arises as a radiative correction if gravitational torsion couples to fermions.
We compute the transport equations for polarized photons from a Boltzmann-like equation, showing
that such a coupling will source circular polarization of the microwave background. For the
particular coupling considered here, the circular polarization effect is always negligible compared
to the rotation of the linear polarization orientation, also derived using the same formalism. We 
note the possibility that limits on microwave background circular polarization may probe other
photon interactions and related fundamental effects such as violations of Lorentz invariance. 
\end{abstract}

\pacs{12.60.Cn, 98.80.-k}
\maketitle

\section{Introduction}
\label{sec:I Intro}

One of the great successes of the standard cosmology is the prediction and measurement of the temperature anisotropies in the cosmic microwave background radiation. Most of these photons have freely propagated since the epoch of last scattering roughly 14 billion years ago and encode the initial
conditions for structure formation.  Measurements are now consistent to high precision with the simplest cosmological models with an initial power-law spectrum of adiabatic perturbations. Linear polarization of the microwave background fluctuations is also a generic result of these models; recent detections of the linear polarization power spectrum and of the cross-correlation between linear polarization and temperature are also consistent with the same cosmological models. 

In general, radiation can have linear polarization, with two degrees of freedom (a polarization
amplitude and orientation) as well as circular polarization, with a single degree of freedom.
It is well known that if an initially unpolarized photon field evolves solely via Compton scattering 
from free electrons plus free streaming, the resulting radiation field can have linear but not
circular polarization. In the tight-coupling regime prior to last scattering when Compton scattering is rapid compared to the cosmological expansion time scale, the cosmic radiation field will be unpolarized.
As the universe cools and the free electrons become bound into neutral hydrogen,
a small linear polarization is generated from the balance of free-streaming and Compton scattering 
during this recombination process, but the resulting microwave background 
radiation today has circular polarization which is identically zero. In this paper, we consider a
generic class of interactions between photons and an external field which can produce
circular polarization. The interactions have been considered in other contexts and are general
enough to be expected in broad classes of theories beyond the standard model of particle physics. 
The same interaction can also arise if non-zero spacetime 
torsion impacts the microwave background radiation.
The goal of this paper is two-fold.  First, we provide an explicit calculation showing how circular polarization can be generically sourced in the microwave background, with the relevant evolution 
equations.  Second, we demonstrate what the underlying microphysics might look like.  

Consider the following extension of the photon sector of quantum electrodynamics:
\begin{eqnarray}
\nonumber \mathcal{L}' &=&\mathcal{L}_{\textsc{maxwell}}+\mathcal{L}_T  \\
			&\equiv&-\frac{1}{4}F_{\mu \nu}F^{\mu\nu} +g\, \epsilon^{\mu \nu \alpha \beta}A_{\mu}T_{\nu}F_{\alpha\beta} 
\label{L-density}
\end{eqnarray}
where $\mathcal{L}_T$ is CPT odd and violates Lorentz invariance and $g$ is the coupling constant of the interaction. Several authors have investigated such a Lorentz-invariance violating extension of QED for a constant 4-vector $T^{\mu}$ (see {\emph{e.g.~}\cite{colladay:1997:lve, colladay:1998:lve2, kostelecky:2002:rlve} and references therein). These so-called \textit{Standard Model Extensions} have been shown to be renormalizable while maintaining gauge invariance \cite{kostelecky:2003:vps}. We consider here only the flat-spacetime interaction term $\mathcal{L}_T$ for simplicity; this will be a good approximation in any cosmological context (see Refs.~\cite{kostelecky:2004:glvsm} and \cite{ kant:2005:mcs} for the curved spacetime generalization, which includes an extra factor of the square root of the metric determinant). If $T_{\mu}$ is fixed as a constant, the Lagrangian density in Eq.~(\ref{L-density}) is $U(1)$ gauge-invariant apart from a boundary term; therefore in these cases the action is gauge invariant. 

It is well known that such an extension should result in optical activity in the propagation of electromagnetic radiation \cite{carroll:1990:pved, lue:1999:mq, balaji:2003:sdcmb, lepora:1998:cbmb}; specifically, a modification to the dispersion relations of free electromagnetic radiation results in a rotation of the plane of linear polarization during propagation. We make no explicit assumption about whether $T_\mu$ is spacelike or timelike, although the timelike case appears
pathological since it leads to a violation of causality and unitarity  \cite{adam:2001:acs} . The magnitude of optical activity of electromagnetic radiation has been constrained by analysis of observational data from cosmological sources and from the microwave background radiation \cite{carroll:1990:pved, carroll:1997:cap, kostelecky:2001:clve, cabella:2007:cptw, feng:2006:a, kahniashvili:2008:a, xia:2008:a, xia:2008:b, komatsu:2009:a, kahniashvili:2009:b}. 
Such a term may arise as a radiative correction following the coupling of gravitational torsion with fermionic matter \cite{desabbata:1980:mercs, desabbata:1981:gismc}. The same term has also been associated with the cancellation of gauge anomalies in QED when the background field $T_{\mu}$ is allowed to couple to the axial current (see \emph{e.g.~}\cite{harvey:2008:wzw}). 

In this paper, we show that in addition to the well-known polarization rotation, such a term may also generate circular polarization, although for the specific case of Eq.~(\ref{L-density}) the circular polarization
is always negligible compared to the polarization rotation. The generation of circular polarization following the optical activity produced 
by $T_\mu$ parallels the Faraday conversion and Faraday rotation effects for propagation in magnetized plasmas; for a discussion in the context of the microwave background, see \cite{cooray:2003:cmbcp}. The observation of circularly polarized microwave background radiation could be
evidence of Lorentz-invariance violation and thus physics beyond the standard model; conversely,
limits on circular polarization may constrain a certain class of standard model extensions. For a related
analysis using an axion-like pseudoscalar coupling to the electromagnetic field, see Ref.~\cite{finelli08}, who also find a nonzero circular polarization and rotation of linear polarization.

In Sec.~\ref{sec:II}, we review the usual description of polarized electromagnetic radiation in terms of Stokes parameters; linear polarization is described by the $Q$ and $U$ parameters, while circular polarization is described by a non-zero $V$ parameter. Section \ref{sec:III} reviews the construction of the Boltzmann-type equation for the photon number density, starting from the quantum-mechanical evolution of the photon density matrix. In Secs.~\ref{sec:IV} and \ref{sec:V}, we calculate the evolution of the Stokes parameters to first and second order, respectively, in the interaction term $\mathcal{L}_T$,
deriving the evolution equation for the $V$ polarization, which is generated from
linear polarization due to the interaction term. We conclude in Sec.~\ref{sec:VI} with estimates of the size of the
$V$ polarization in the microwave background for given interactions along with the
magnitude of linear polarization rotation. The mathematical details of evaluating the first and second-order interaction terms are relegated to Appendices A and B, while Appendix C addresses
the issue of gauge invariance.

\section{Stokes Parameters}
\label{sec:II}

The polarization state of light is most easily described by making use of the Stokes parameters. For a complete review see, {\emph{e.g.}, Refs.~\cite{mcmaster:1954:psp,mcmaster:1961:mrp,jackson:1999:ced} or any optics text. Here we review the basic construction of the Stokes parameters in the classical and quantum mechanical contexts in order to motivate the  quantum field theoretical construction. Consider a classical electromagnetic plane wave with electric field given by the  components
\begin{eqnarray}
	\label{eq:2.2}
		E _{1}(t)=a_{1}\sin(\omega t- \epsilon _1)	\quad \text{and} \quad 	E _{2}(t)=a_{2}\sin(\omega t- \epsilon _2)		
\end{eqnarray} 
where we assume, for simplicity, that the wave is nearly monochromatic with frequency $\omega$, such that $a_1$, $a_2$,  $\epsilon _1$, and  $\epsilon _2$ only vary on time scales long
compared to $\omega^{-1}$. The Stokes parameters in the linear polarization basis are then defined as
\begin{eqnarray}
	\label{eq:2.3}
		I &\equiv & \left\langle(a_{1})^2 + (a_{2})^2\right\rangle, \\
	\label{eq:2.4}		
		Q &\equiv & \left\langle(a_{1})^2 - (a_{2})^2\right\rangle,\\
	\label{eq:2.5}	
		U &\equiv & \left\langle2a_{1}a_{2}\,\cos\delta\right\rangle,\\
	\label{eq:2.6}	
		V &\equiv & \left\langle2a_{1}a_{2}\,\sin\delta\right\rangle,
\end{eqnarray} 
where  $\delta \equiv \epsilon _2-\epsilon _1$ and the brackets signify a time
average over a time long compared to $\omega^{-1}$. The $I$ parameter measures the intensity of the radiation, while the parameters $Q$, $U$, and $V$ each carry information about the polarization of the radiation. Unpolarized radiation is described by $Q = U = V = 0$. The linear polarization of the radiation is encoded in $Q$ and $U$, while the parameter $V$ 
is a measure of elliptical polarization with the special case of circular polarization ocurring when $a_1 = a_2$ and $\delta = \pm \pi/2$. From here on we will simply refer to $V$ as the measure of circular polarization, which is technically correct if $Q=0$. Note that while $I$ and $V$ are coordinate independent, $Q$ and $U$ depend on the orientation of the coordinate system used on the plane orthogonal to the direction of propagation. Under a rotation of the coordinate system by an angle $\phi$, the parameters $Q$ and $U$ transform according to
\begin{eqnarray}
\nonumber Q' &=& Q\, \cos(2\phi )+ U \, \sin(2\phi ), \\
\nonumber U' &=& -Q\, \sin (2\phi ) + U\,\cos(2\phi ),
\end{eqnarray}
while the angle defined by
\begin{equation}
\nonumber \Phi = \half \arctan \bigg( \frac{U}{Q} \bigg)
\end{equation}
goes to $\Phi - \phi$ following a rotation by the angle $\phi$. Therefore, $Q$ and $U$ only define an orientation and not a particular direction in the plane: after a rotation by $\pi$ they are left unchanged. Physically, this is simply a manifestation of the oscillatory behavior of the electric field. These properties indicate that $Q$ and $U$ are part of a second-rank symmetric trace-free tensor $P_{ab}$, i.e. a spin-2 field in the plane orthogonal to the direction of propagation. Such a tensor can be represented as
\begin{equation}
\label{eq:2.7}
 P_{ij} 
=\left(
\begin{array}{cc}
P & 0  \\
0 & -P \\  \end{array}
\right), 
\end{equation}
in an orthonormal eigenbasis, where $P = (Q^2+U^2)^{1/2}$ is often called the magnitude of linear polarization.

In quantum mechanics we can express the state of a photon $\mathcal{A}$ as    
\begin{eqnarray}
\label{eq:2.8}
\ket{\mathcal{A}} =  \displaystyle\sum_{i}a_i\ket{\epsilon_i},
\end{eqnarray}
where $\ket{\epsilon_i}$ ($i=1,2$) span the polarization state space and $a_i$ are in general complex. The projection operators
\begin{eqnarray}
\label{eq:2.9}
		\hat I &\equiv & \ket{\epsilon_1}\bra{\epsilon_1} +\ket{\epsilon_2}\bra{\epsilon_2}, \\
	\label{eq:2.10}		
		\hat Q &\equiv & \ket{\epsilon_1}\bra{\epsilon_1} -\ket{\epsilon_2}\bra{\epsilon_2},\\
	\label{eq:2.11}	
		\hat U &\equiv & \ket{\epsilon_1}\bra{\epsilon_2} +\ket{\epsilon_2}\bra{\epsilon_1},\\
	\label{eq:2.12}	
		\hat V &\equiv & i\ket{\epsilon_2}\bra{\epsilon_1} -i\ket{\epsilon_1}\bra{\epsilon_2}
\end{eqnarray}
have expectation values in single photon states which give the classical Stokes parameters, 
Eqs.~(\ref{eq:2.3})--(\ref{eq:2.6}). In a general mixed state, the density matrix $\rho$
on the polarization state space encodes the intensity and polarization of the photon ensemble. For example,
\begin{equation}
\nonumber \langle \hat Q \rangle = \frac{{\rm tr}(\rho \hat Q)}{{\rm tr}(\rho)}=\frac{1}{{\rm tr}(\rho )}{\rm tr}\bigg[ \left( \begin{array}{cc}\rho_{11} & \rho_{12}  \\ \rho_{21} & \rho_{22} \\  \end{array}\right) \left( \begin{array}{cc}1 & 0  \\ 0 & -1 \\  \end{array}\right) \bigg ] = \frac{\rho_{11}-\rho_{22} }{{\rm tr}(\rho )} .
\end{equation}
Similar relations hold for the other ``Stokes operators" such that the density matrix can be represented as
\begin{eqnarray}
\label{eq:2.13}
 \rho 
=\frac{{\rm tr}(\rho)}{2} \left(
\begin{array}{cc}
1+Q & U-iV  \\
U+iV & 1-Q \\  
\end{array}
\right), 
\end{eqnarray}
where $Q= \langle \hat Q \rangle$, $U= \langle \hat U \rangle$, and $V= \langle \hat V \rangle$.

\section{The Photon Boltzmann Equation}
\label{sec:III}
We now review the construction of the evolution equation for the photon number operator under the influence of some perturbation to a free theory. The following formalism was developed to study neutrino mixing and damping \cite{raffelt:1993:nabe, raffelt:1993:gkmn}. It has also been applied to describe the generation of linear polarization in the microwave background due to Compton scattering during recombination \cite{kosowsky:1996:cmbp}, generalizing an earlier
kinetic equation treatment of microwave background temperature fluctuations \cite{dod95}. 

Consider an ensemble of free photons. We will assume that the interaction $\mathcal{L}_T$ is slowly ``turned on" and that the interactions of the photons with the external field $T_{\mu}$ are localized such that the photons can be considered free (with respect to the interaction $\mathcal{L}_T$) both before and after each point interaction. That is, we make the usual assumptions of scattering theory. We will not consider any possible interference effects which might occur between $\mathcal{L}_T$ and any other interaction.

The free photon field in the Coulomb (radiation) gauge can be expressed as
\begin{equation}
	\label{eq:3.14}
		\hat A_\mu(x) = \int \frac{d^3 {\bf k}}{(2\pi)^3 2 k^0} \left[ \hat a_s(k) \epsilon _{s\mu}(k)
		e^{-ik\cdot x}+\hat a_s^\dagger (k) \epsilon^* _{s\mu}(k)e^{ik\cdot x}  \right],
\end{equation}
where $\epsilon _{s\mu}(k)$ are the photon polarization 4-vectors and $s$, which takes the
values 1 and 2, indexes the orthogonal transverse polarizations. The free creation and annihilation operators satisfy the canonical commutation relation
\begin{equation}
	\label{eq:3.15}
		\left[ \hat a_i (k),\hat a_j^\dagger (k')\right] = (2\pi )^3 2k^0\delta_{ij}\delta^{(3)}({\bf k} - {\bf k}' )
\end{equation}
where $k^0= \left|{\bf k}\right|$.

We will be interested in the evolution of the polarization state of a photon ensemble, which 
is completely characterized by the density matrix $\rho_{ij}$ defined via 
\begin{equation}
	\label{eq:3.16}
		\langle \hat a_i^\dagger (k)\hat a_j (k')\rangle = (2\pi )^3 2k^0\delta^{(3)}({\bf k} - {\bf k}' )\rho_{ij}({\bf k}).
\end{equation}
The number operator $ \mathcal{\hat D}_{ij}({\bf k}) = \hat a_i^\dagger({\bf k})\hat a_j ({\bf k})$, according to (\ref{eq:3.16}), is related to $\rho_{ij}({\bf k})$ as 
\begin{equation}
	\label{eq:3.17}
		\langle \mathcal{\hat D}_{ij}({\bf k}) \rangle = (2\pi )^3 2k^0\delta^{(3)}(0)\rho_{ij}({\bf k}).
\end{equation}
The infinite factor $\delta^{(3)}(0)$ is a remnant of the infinite quantization volume. As we show explicitly below, it cancels from all final expressions. Motivated by the construction in the quantum mechanical system above, we can project out quantities analogous to the classical Stokes parameters:
\begin{eqnarray}
	\label{eq:3.18}
	{\rm tr}(\sigma_0 \rho({\bf k}))&\rightarrow & I \propto \langle \mathcal{\hat H}_0({\bf k}) \rangle \\
	\label{eq:3.19}
	{\rm tr}(\sigma_z \rho ({\bf k}))&\rightarrow &Q \propto \langle \mathcal{\hat D}_{11}({\bf k}) \rangle -\langle \mathcal{\hat D}_{22}({\bf k}) \rangle \\ 
	\label{eq:3.20}
	{\rm tr}(\sigma_x \rho ({\bf k}))&\rightarrow &U \propto \langle \mathcal{\hat D}_{12}({\bf k}) \rangle +\langle \mathcal{\hat D}_{21}({\bf k}) \rangle \\
	\label{eq:3.21}
	{\rm tr}(\sigma_y \rho ({\bf k}))&\rightarrow &iV \propto \langle \mathcal{\hat D}_{12}({\bf k}) \rangle -\langle \mathcal{\hat D}_{21}({\bf k}) \rangle 
\end{eqnarray}
where $\sigma_0$ is the $2\times 2$ identity matrix, $\sigma_i$ are the Pauli matrices and the trace is over polarization indices. $\mathcal{\hat H}_0({\bf k})$ is the free energy density operator of the mode with wavenumber ${\bf k}$.

We ignore any correlations such as $\langle \hat a_i (k)\hat a_j (k')\rangle$ and $\langle \hat a_i^\dagger (k) \hat a_j^\dagger (k')\rangle$ which might be generated via the interaction $\mathcal{L}_T$. In essence we are assuming that the background field $T_{\mu}$ varies slowly enough in time so that physical two-photon states are neither created nor destroyed by the interaction $\mathcal{L}_T$. If we define $\omega_T$ as a characteristic energy scale of the background field $T_{\mu}$ 
and $\omega$ as the frequency of a particular mode associated with free oscillations of the creation and annihilation operators, then we are requiring that $\omega_T\ll \delta\omega/\omega\ll\omega$ \cite{raffelt:1992:qspm}. Note that $\delta\omega/\omega$ is the order at which mixing occurs between $\hat a_i$ and $\hat a_i^\dagger$, and such a mixing will result in a variation $\delta\langle \hat a_i (k)\hat a_j (k')\rangle \simeq (\delta\omega/\omega)\langle \hat a_i^\dagger (k)\hat a_j (k')\rangle$, which again we neglect. Naively, we then expect the following formalism to fail for low-frequency photon modes, although at precisely what scale the approximation breaks down 
depends on the characteristic scale of the background field $T_{\mu}$.

The evolution of the number operator $ \mathcal{\hat D}_{ij}({\bf k})$ can be computed using
a perturbative expansion in the interaction strength. The following reviews 
the construction detailed in \cite{raffelt:1993:gkmn}. Recall that the time evolution of any Heisenberg picture operator is given by
\begin{equation}
	\label{eq:3.22}
		\frac{d}{dt} \mathcal{\hat D}_{ij} = i\big [ \hat H,\mathcal{\hat D}_{ij}].
\end{equation}
If the full Hamiltonian can be split into a free and interacting part, $\hat H = \hat H_{\text{free}}+\hat H_{\text{int}}$, then (\ref{eq:3.22}) becomes.
\begin{equation}
	\label{eq:3.23}
		\frac{d}{dt} \mathcal{\hat D}_{ij} = i\big [ \hat H_{\text{free}},\mathcal{\hat D}_{ij}] + i\big [ \hat H_{\text{int}},\mathcal{\hat D}_{ij}].
\end{equation}
A first order perturbative approximation for the evolution of the number operator is given by replacing all operators on the right hand side of equation (\ref{eq:3.23}) by their free theory counterparts, e.g. 
\begin{equation}
	\label{eq:3.24}
		\left\langle \frac{d}{dt}\mathcal{\hat D}_{ij} \right\rangle \simeq i\langle [ \hat H^0_{\text{int}} , \mathcal{\hat D}^0 ] \rangle,
\end{equation}
where $\mathcal{\hat O}^0$ corresponds to the operator $\mathcal{\hat O}$ evaluated in terms of the operators of the free theory. The above assumes that $[ \hat H^0_{\text{free}} , \mathcal{\hat D}^0 ]=0$. We will refer to the term on the right hand side of (\ref{eq:3.24}) as the refractive term, or the forward scattering term. To determine a second order perturbative approximation, we will use the fact that we can expand any operator to first order in interactions as
		\begin{equation}
		\label{eq:3.25}
			\hat \xi (t) \simeq \hat \xi^0 (t)+i\int_0^t dt' [\hat H^0_{\text{int}}(t- t'),\hat \xi^0 (t)],
		\end{equation}
with the initial conditions $\hat \xi (0)=\hat \xi^0 (0)$. The expansion Eq.~(\ref{eq:3.25}) can be verified by explicitly taking the time derivative of both sides and seeing that one recovers the Heisenberg equation to first order in interactions. We now expand $[ \hat H_{\text{int}} , \mathcal{\hat D} ] $ as in (\ref{eq:3.25}), insert the result into (\ref{eq:3.23}), and upon the evaluation of all operators in terms of the free theory operators arrive at
\begin{equation}
\label{eq:3.26}
		\left\langle \frac{d}{dt}\mathcal{\hat D}_{ij} \right\rangle(t) \simeq i\langle [ \hat H^0_{\text{int}}(t) , \mathcal{\hat D}^0 ] \rangle -\int_0^t dt' \langle [\hat H^0_{\text{int}}(t-t'),[ \hat H^0_{\text{int}}(t) , \mathcal{\hat D}^0 ] ]\rangle . 
\end{equation}
The second term in (\ref{eq:3.26}) will be referred to as the damping term or the non-forward scattering term. In terms of $\rho$, the evolution equation reads
\begin{equation}
\label{eq:3.27}
		(2\pi )^3 \delta^3 (0)2q^0 \frac{d} {dt} \rho_{ij}(0,{\bf q})=i\langle[\hat H_{\text{int}}^0(0),\mathcal{\hat D}^0_{ij}({\bf q})]\rangle -  \frac{1}{2}\int_{-\infty}^\infty dt'\phantom{1}\langle[\hat H_{\text{int}}^0(t'),[\hat H_{\text{int}}^0(0),\mathcal{\hat D}^0_{ij}({\bf q})]]\rangle,
\end{equation}
where, as mentioned above, all factors of $\delta^3(0)$ will cancel from the final expressions. In going from (\ref{eq:3.26}) to (\ref{eq:3.27}), we have assumed the time step $t$ in Eq.~(\ref{eq:3.26}) is both small relative to the characteristic time scale of the evolution of $\rho$ and large relative to the duration of a single interaction. This allows us to take $t\rightarrow \infty$ and set $\rho(t)=\rho(0)$ \cite{raffelt:1993:gkmn}. We have then replaced the integral $(\int_0^\infty dt)$ with $(\frac{1}{2}\int_{-\infty}^\infty dt)$, the difference being a principle part integral which is a second order correction to the refractive term. Equation~(\ref{eq:3.27}) can be viewed as a \textit{Generalized Boltzmann Equation} for the phase space function $\rho$. In this approximation, we have a set of differential equations for the components $\rho_{ij}$ at $t=0$; if the interactions are ``forgotten'' between intermediate collisions (an assumption known as molecular chaos in the derivation of the standard Boltzmann equation), then the differential equations will be valid for all times over which the interaction is relevant. 
The Liouville terms on the left side will incorporate any effects which result from a departure of the spacetime metric from a flat metric, including any weak inhomogeneities due to the presence of gravitational perturbations about a homogeneous cosmology. The case $H_{\text{int}}(t) = \int d{\bf x} \bar \psi \gamma^{\mu}A_\mu \psi$ as in full QED, where $\psi$ is a spinor field associated with the electron, recovers the radiative transfer equations of Chandrasekhar \cite{chandrasekhar:1960:rt} in the appropriate limits (see \cite{kosowsky:1996:cmbp} for more details).

As we will see below, this construction allows linearizing the right side of Eq.~(\ref{eq:3.27}) 
in $\rho$. We can expand the photon density matrix about a uniform unpolarized distribution 
(ignoring any small inhomogeneities) as
\begin{equation}
	\label{eq:3.28}
		\rho_{ij}(t, k, \hat k) =\rho^{(0)}_{ij}(t, k) +\rho^{(1)}_{ij}(t, k, \hat k), 
\end{equation}
where $\rho^{(0)}_{11}=\rho^{(0)}_{22}$ and $\rho^{(0)}_{12}=\rho^{(0)}_{21}=0$. As a consistency check, the right side of Eq.~(\ref{eq:3.27}) should vanish when evaluated in terms of $\rho^{(0)}$ so that in for example an FRW background of zero spatial curvature with scale factor $a(t)$ we have \cite{kosowsky:1996:cmbp}
\begin{equation}
	\label{eq:3.29}
		\frac{d}{dt}\rho^{(0)}_{11} = \frac{\partial\rho^{(0)}_{11}}{\partial t}-\frac{\dot a}{a}k
		\frac{\partial\rho^{(0)}_{11}}{\partial k}= 0,
\end{equation}
the solution of which is $\rho^{(0)}_{11}(t,k)= \rho^{(0)}_{11}(ka)$, recovering the uniform redshift due to cosmological expansion.

In order to make contact with the measurable Stokes parameters, we define the normalized brightness perturbations \cite{kosowsky:1996:cmbp}
\begin{eqnarray}
	\label{eq:3.30}
		\Delta _I &\equiv & \Big[\frac{q}{4}\frac{\partial\rho_{11}^{(0)}(q)}{\partial q}\Big]^{-1} ( \rho^{(1)}_{11}+\rho^{(1)}_{22}) ,\\
	\label{eq:3.31}
		\Delta _Q &\equiv &\Big[\frac{q}{4}\frac{\partial\rho_{11}^{(0)}(q)}{\partial q}\Big]^{-1} ( \rho^{(1)}_{11}-\rho^{(1)}_{22}),\\
	\label{eq:3.32}
		\Delta _U &\equiv &\Big[\frac{q}{4}\frac{\partial\rho_{11}^{(0)}(q)}{\partial q}\Big]^{-1} ( \rho^{(1)}_{12}+\rho^{(1)}_{21}),\\
	\label{eq:3.33}
		\Delta _V &\equiv &-i\Big[\frac{q}{4}\frac{\partial\rho_{11}^{(0)}(q)}{\partial q}\Big]^{-1} ( \rho^{(1)}_{12}-\rho^{(1)}_{21}),
\end{eqnarray} 
where $q=ka$ is the comoving photon momentum and we have expanded the density matrix 
$\rho$ in a linear polarization basis.
		

\section{The First-Order Interaction Term} 
\label{sec:IV} 

We now evaluate the right side of the
Generalized Boltzmann Equation (\ref{eq:3.27}) 
for an interaction Hamiltonian which is linear in the Hamiltonian density
\begin{equation}
	\label{eq:4.34}
		\mathcal{\hat H}_{T} = -g \epsilon^{\mu\nu\alpha\beta}:\hat A_{\mu} T_{\nu} \hat F_{\alpha \beta}: \, ,
\end{equation}
where $\hat A_\mu$ for the free theory is given by Eq.~(\ref{eq:3.14}) and $\hat F_{\mu \nu} = 2\partial_{[\mu} \hat A _{\nu]}$ is the free electromagnetic field strength operator. We will treat $T_\mu$ as a classical background field, the dynamics of which are not influenced by the electromagnetic interaction Eq.~(\ref{eq:4.34}) and are assumed to be externally prescribed. The symbol $:\,\cdots\,:$ denotes normal ordering of the enclosed operator products.

Following a local $U(1)$ gauge transformation $\delta_G A_\mu = \partial_\mu\lambda$ of $A_\mu$, 
the resulting change in the Hamiltonian density Eq.~(\ref{eq:4.34}) is given by
\begin{eqnarray}
\label{eq:4.35}
\delta_G\mathcal{H}_{T} &=& -g \epsilon^{\mu\nu\alpha\beta} (\partial_\mu\lambda) T_{\nu} F_{\alpha \beta} = g \epsilon^{\mu\nu\alpha\beta} \lambda (\partial_{[ \mu} T_{\nu ]}) F_{\alpha \beta} - g \epsilon^{\mu\nu\alpha\beta}\partial_\mu (\lambda T_{\nu} F_{\alpha \beta}),
\end{eqnarray}
where the square brackets denote anti-symmetrization and we have used $\partial_{[\mu}F_{\alpha\beta]}=0$. If $\partial_{[\mu} T_{\nu]}=0$ then the interaction Hamiltonian is gauge invariant apart from a boundary term. We assume that the external field $T_\mu(x)$ does indeed satisfy this condition allowing us to maintain gauge invariance, which is detailed in the second-order calculation in Appendix~\ref{sec-C}. Furthermore, we assume that $T_\mu(x)$ is a pseudo-vector: under a parity transformation the external field transforms as 
\begin{eqnarray}
\label{eq:4.36}
T_\mu (t,-{\bf x}) =\bigg\{ 
\begin{array}{ll}
  -T_\mu (t,{\bf x}),& \mu = 0  \\
 +T_\mu (t,{\bf x}), & \mu =1,2,3  \\
\end{array}
\end{eqnarray}
and under time reversal the external field transforms as
\begin{eqnarray}
\label{eq:4.37}
T_\mu (-t,{\bf x}) =\bigg\{ 
\begin{array}{ll}
  +T_\mu (t,{\bf x}),& \mu = 0  \\
 -T_\mu (t,{\bf x}), & \mu =1,2,3 . \\
\end{array}
\end{eqnarray}
We will find it useful to consider the Fourier transform of $T_\mu$,
\begin{equation}
	\label{eq:4.38}
		\tilde T _\mu (p) = \int d^4 x \phantom{1}T_\mu(x) e^{-i p \cdot x} .
\end{equation}
In momentum space the gauge invariance restrictions can be expressed by the condition
\begin{equation}
	\label{eq:4.39}
		p_{[\mu}\tilde T _{\nu]} (p) = 0 .
\end{equation}
In order to determine whether circular polarization can be sourced at some order of (\ref{eq:4.35}), we will not need to impose any further restrictions on  $T_\mu(x)$ aside from those listed above. Specifically, our calculations do not assume that $T_\mu(x)$ is either timelike or spacelike. As indicated by the investigation in \cite{adam:2001:acs}, violations of causality and unstable solutions may arise for the case of timelike $T_\mu(x)$, and the results here must be interpreted with care in this case. 

In Appendix~\ref{sec-A} below we detail the calculation of the first-order interaction Hamiltonian which is linear in the Hamiltonian density Eq.~(\ref{eq:4.34}), as well as the refractive and damping terms of Eq.~(\ref{eq:3.27}) due to first-order processes. Quoting the results of Appendix~\ref{sec-A}, the interaction Hamiltonian for first-order processes is given by
\begin{equation}
	\label{eq:4.40}
		\phantom{}^{1} \hat H_{\text{int}}(t) = -2 ig\int dp\,d{\bf k}\, \frac{\tilde T_\nu(p) e^{i(k_0+p_0-\left|{\bf k}+{\bf p}\right|)t}}{2\left|{\bf k}+{\bf p}\right|} \hat a_s^\dagger (k)\hat a_r (\tilde k) \epsilon^* _{s\mu}({\bf k})\big(\epsilon^{\mu\nu 0\beta}(\left|{\bf k}\right|+\left|{\bf k}+{\bf p}\right|)+\epsilon^{\mu\nu j\beta}(2{k}_j+{p}_j) \big)\epsilon _{r\beta}({\bf k}+{\bf p} ).
\end{equation}
The refractive term of Eq.~(\ref{eq:3.27}) for this first order interaction Hamiltonian is given by 
\begin{eqnarray}
	\label{eq:4.41}
		i\left\langle\left[ \phantom{}^{1}\hat H_{\text{int}}(0),\mathcal{\hat D}_{uv}({\bf q})\right]\right\rangle &=&4g q^0(2\pi )^3 \delta^3(0) \epsilon _{s\mu}({\bf q})\left( \delta_{ur}\rho_{sv}( {\bf q})-\delta_{vs}\rho_{ur}({\bf q})\right)\mathcal{ A}^{\mu \beta}(q) \epsilon _{r\beta}({\bf q}) ,
\end{eqnarray}
where $\mathcal{A}^{\mu\beta}$ is defined via 
\begin{eqnarray}
	\label{eq:4.42}
		(2\pi )^3\delta^3(0)\mathcal{ A}^{\mu \beta}(q) =(2\pi )^3\delta^3(0)\mathcal{ A}^{[\mu \beta]}(q)= \int \frac{ dp}{2\left|{\bf q}+{\bf p}\right|} (2\pi )^3\delta^{3}(-{\bf p} )\tilde T_\nu(p)\epsilon^{\mu\nu \alpha\beta}(2{q}_\alpha+\tilde p_\alpha) ,
\end{eqnarray}
and we have for convenience defined $\tilde p \equiv  (\Delta q, {\bf p})$, $\Delta q \equiv \left|{\bf q}+{\bf p}\right|-\left|{\bf q}\right|$.

As detailed in Appendix~\ref{sec-A} the damping term of Eq.~(\ref{eq:3.27}) due to first order processes is given by
\begin{eqnarray}
	\label{eq:4.43}
\nonumber  D_{uv} &\equiv & \int_{-\infty}^\infty dt'\left\langle\left[ \phantom{}^{1}\hat H_{\text{int}}(t'),\left[ \phantom{}^{1}\hat H_{\text{int}}(0),\mathcal{\hat D}_{uv}({\bf q})\right]\right]\right\rangle \\
& =&  -(2ig)^2 \epsilon _{m\kappa}\epsilon _{s\mu}  \int  \frac{dp }{2\left|{\bf q}+{\bf p}\right| } \tilde T_\lambda(\tilde p)q_\rho\epsilon^{\sigma\lambda\rho\kappa} \tilde T_\nu(p) (2{  q}+{ \tilde p})_\alpha \epsilon^{\mu\nu \alpha\beta}\tilde\epsilon _{n\sigma}\tilde\epsilon _{r\beta} \\
\nonumber	&&\qquad\qquad\qquad\times\bigg(  \delta_{nr} \delta_{us}  \rho_{mv}({\bf q}) + \delta_{rn}\delta_{vs}\rho_{um}({\bf q})-\delta_{us} \delta_{mv} \rho_{rn}({\bf q}+{\bf p})-\delta_{um} \delta_{sv} \rho_{nr}({\bf q}+{\bf p}) \bigg) ,
\end{eqnarray}
where we have defined $\epsilon_{r\mu}\equiv\epsilon_{r\mu}({\bf q})$ and $\tilde\epsilon_{r\mu}\equiv\epsilon_{r\mu}({\bf q}+{\bf p})$.

Once the density matrix is expanded as given in Eq.~(\ref{eq:3.28}), it is straightforward to see that the refractive term, Eq.(\ref{eq:4.41}), vanishes when evaluated in terms of $\rho^{(0)}(\left | {\bf q} \right|)$. For the damping term, Eq.~(\ref{eq:4.43}), we must perform an expansion of $\rho^{(0)}(\left | {\bf q} + {\bf p} \right|)=\rho^{(0)}(\left | {\bf k} \right|)$, in the $p$ integral of Eq.~(\ref{eq:4.43}), about $\left | {\bf q} \right|$:
\begin{eqnarray}
\label{eq:4.44}
\rho^{(0)}(\left | {\bf k} \right|) = \rho^{(0)}(\left| {\bf q} \right| ) +\mathcal{O}\bigg(\frac{d\rho^{(0)}}{d\left| {\bf k} \right|}(\left| {\bf q} \right|)\bigg),
\end{eqnarray}
which is a suitable approximation as long as $\tilde T(p^0,{\bf p})$ has support solely over $\left|{\bf p}\right|\ll \left|{\bf q}\right|$, where $\left|{\bf q}\right|$ is the energy of the scattering photons. Then to lowest order in $\rho^{(0)}$ the damping term vanishes and we have
\begin{eqnarray}
\label{eq:4.45}
\frac{d} {dt} \rho^{(0)}=  0 + \mathcal{O}\bigg(g^2\frac{d\rho^{(0)}(\left| {\bf q} \right|)}{d\left| {\bf q} \right|}\bigg) .
\end{eqnarray}

In terms of the Stokes brightness perturbations defined in Eqs.~(\ref{eq:3.30})--(\ref{eq:3.33}), to first order in $g$
the evolution of the polarization of the photon ensemble becomes
\begin{eqnarray}
	\label{eq:4.46}
		\frac{d} {dt} \Delta _I &=& 0, \\
	\label{eq:4.47}
		\frac{d} {dt} \Delta _Q &=& -g\alpha (q) \Delta _U , \\
	\label{eq:4.48}
		\frac{d} {dt} \Delta _U &=& g\alpha (q)\Delta _Q , \\
	\label{eq:4.49}
		\frac{d} {dt} \Delta _V &=& 0,
\end{eqnarray}
where we have defined the quantity 
\begin{eqnarray}
\label{eq:4.50}
	\alpha (q) \equiv 4 \epsilon _{1\mu}(q)\mathcal{A}^{\mu\beta} (q) \epsilon _{2\beta}(q).
\end{eqnarray}

For processes which are first order in the Hamiltonian density Eq.~(\ref{eq:4.34}), according to 
Eq.~(\ref{eq:4.49}) no circular polarization is generated to $\mathcal{O}(g)$ in our approximation. In fact, it is easy to see that to $\mathcal{O}(g)$ (the refractive term), Eqs.~(\ref{eq:4.46})--(\ref{eq:4.49}) reproduce the well-known effect of optical activity of the electromagnetic radiation, rotating the plane of linear polarization during propagation \cite{carroll:1990:pved, lue:1999:mq, balaji:2003:sdcmb,lepora:1998:cbmb} . 
This is a useful check of the calculations.


The relevant linear combinations of the damping term Eq.(\ref{eq:4.43}) which
source the polarization of the photon ensemble to $\mathcal{O}(g^2)$ and due to first order processes are
\begin{eqnarray*}
D_{11}+ D_{22}&=& -(2ig)^2  \int  \frac{dp }{2\left|{\bf q}+{\bf p}\right| } \tilde T_\lambda(\tilde p)q_\rho\epsilon^{\sigma\lambda\rho\kappa} \tilde T_\nu(p) (2{  q}+{ \tilde p})_\alpha \\
&&\times\bigg\{(\tilde\epsilon _{1\sigma} \tilde\epsilon _{1\beta}+\tilde\epsilon _{2\sigma} \tilde\epsilon _{2\beta}) \bigg[ 2\epsilon _{1\kappa}\epsilon _{1\mu}  \rho^{(1)}_{11}({\bf q})+ 2\epsilon _{2\kappa}\epsilon _{2\mu}  \rho^{(1)}_{22}({\bf q})+ (\epsilon _{2\kappa}\epsilon _{1\mu}+ \epsilon _{1\kappa}\epsilon _{2\mu}) [ \rho^{(1)}_{12}({\bf q})+ \rho^{(1)}_{21}({\bf q})] \bigg]\\
&&\qquad-\big(\epsilon _{1\kappa}\epsilon _{1\mu}+ \epsilon _{2\kappa}\epsilon _{2\mu}\big)\bigg [ 2\tilde\epsilon _{1\beta}\tilde\epsilon _{1\sigma}\rho^{(1)}_{11}({\bf q}+{\bf p})+2\tilde\epsilon _{2\beta}\tilde\epsilon _{2\sigma}\rho^{(1)}_{22}({\bf q}+{\bf p})\\
&&\qquad\qquad\qquad\qquad\qquad\qquad +(\tilde\epsilon _{1\beta}\tilde\epsilon _{2\sigma}+\tilde\epsilon _{2\beta}\tilde\epsilon _{1\sigma})[\rho^{(1)}_{12}({\bf q}+{\bf p})+\rho^{(1)}_{21}({\bf q}+{\bf p})]\bigg ] \bigg\} +\mathcal{O}\bigg(g^2\frac{d\rho^{(0)}(\left| {\bf q} \right|)}{d\left| {\bf q} \right|}\bigg),
\end{eqnarray*}
\begin{eqnarray*}
D_{11}- D_{22}&=& -(2ig)^2  \int  \frac{dp }{2\left|{\bf q}+{\bf p}\right| } \tilde T_\lambda(\tilde p)q_\rho\epsilon^{\sigma\lambda\rho\kappa} \tilde T_\nu(p) (2{  q}+{ \tilde p})_\alpha \\
&&\times\bigg\{(\tilde\epsilon _{1\sigma} \tilde\epsilon _{1\beta}+\tilde\epsilon _{2\sigma} \tilde\epsilon _{2\beta}) \bigg[ 2\epsilon _{1\kappa}\epsilon _{1\mu}  \rho^{(1)}_{11}({\bf q})- 2\epsilon _{2\kappa}\epsilon _{2\mu}  \rho^{(1)}_{22}({\bf q})+ (\epsilon _{2\kappa}\epsilon _{1\mu}- \epsilon _{1\kappa}\epsilon _{2\mu}) [ \rho^{(1)}_{12}({\bf q})+ \rho^{(1)}_{21}({\bf q})] \bigg]\\
&&\qquad-\big(\epsilon _{1\kappa}\epsilon _{1\mu}- \epsilon _{2\kappa}\epsilon _{2\mu}\big)\bigg [ 2\tilde\epsilon _{1\beta}\tilde\epsilon _{1\sigma}\rho^{(1)}_{11}({\bf q}+{\bf p})+2\tilde\epsilon _{2\beta}\tilde\epsilon _{2\sigma}\rho^{(1)}_{22}({\bf q}+{\bf p})\\
&&\qquad\qquad\qquad\qquad\qquad\qquad+(\tilde\epsilon _{1\beta}\tilde\epsilon _{2\sigma}+\tilde\epsilon _{2\beta}\tilde\epsilon _{1\sigma})[\rho^{(1)}_{12}({\bf q}+{\bf p})+\rho^{(1)}_{21}({\bf q}+{\bf p})]\bigg ] \bigg\} +\mathcal{O}\bigg(g^2\frac{d\rho^{(0)}(\left| {\bf q} \right|)}{d\left| {\bf q} \right|}\bigg),
\end{eqnarray*}
\begin{eqnarray*}
D_{12}+ D_{21}&=& -(2ig)^2  \int  \frac{dp }{2\left|{\bf q}+{\bf p}\right| } \tilde T_\lambda(\tilde p)q_\rho\epsilon^{\sigma\lambda\rho\kappa} \tilde T_\nu(p) (2{  q}+{ \tilde p})_\alpha \\
&&\times\bigg\{(\tilde\epsilon _{1\sigma} \tilde\epsilon _{1\beta}+\tilde\epsilon _{2\sigma} \tilde\epsilon _{2\beta})\bigg[(\epsilon _{1\kappa} \epsilon _{1\mu}+\epsilon _{2\kappa}\epsilon _{2\mu}) [ \rho^{(1)}_{12}({\bf q})+\rho^{(1)}_{21}({\bf q})] +2\epsilon _{1\kappa} \epsilon _{2\mu}\rho^{(1)}_{11}({\bf q})+ 2\epsilon _{2\kappa}\epsilon _{1\mu} \rho^{(1)}_{22}({\bf q})\bigg]\\
&&\qquad-(\epsilon _{2\kappa}\epsilon _{1\mu}+ \epsilon _{1\kappa}\epsilon _{2\mu})\bigg[ 2\tilde\epsilon _{1\beta}\tilde\epsilon _{1\sigma}\rho^{(1)}_{11}({\bf q}+{\bf p})+2\tilde\epsilon _{2\beta}\tilde\epsilon _{2\sigma}\rho^{(1)}_{22}({\bf q}+{\bf p}) \\
&&\qquad\qquad\qquad\qquad\qquad+(\tilde\epsilon _{1\beta}\tilde\epsilon _{2\sigma}+ \tilde\epsilon _{1\sigma}\tilde\epsilon _{2\beta}) [ \rho^{(1)}_{12}({\bf q}+{\bf p})+  \rho^{(1)}_{21}({\bf q}+{\bf p})] \bigg ] \bigg\} +\mathcal{O}\bigg(g^2\frac{d\rho^{(0)}(\left| {\bf q} \right|)}{d\left| {\bf q} \right|}\bigg),
\end{eqnarray*}
\begin{eqnarray*}
D_{12}- D_{21}&=& -(2ig)^2  \int  \frac{dp }{2\left|{\bf q}+{\bf p}\right| } \tilde T_\lambda(\tilde p)q_\rho\epsilon^{\sigma\lambda\rho\kappa} \tilde T_\nu(p) (2{  q}+{ \tilde p})_\alpha \\
&&\qquad\times\bigg\{(\tilde\epsilon _{1\sigma} \tilde\epsilon _{1\beta}+\tilde\epsilon _{2\sigma} \tilde\epsilon _{2\beta})(\epsilon _{1\kappa} \epsilon _{1\mu}+\epsilon _{2\kappa}\epsilon _{2\mu}) [\rho^{(1)}_{12}({\bf q})-\rho^{(1)}_{21}({\bf q})]\\
&&\qquad\qquad-(\epsilon _{2\kappa}\epsilon _{1\mu}- \epsilon _{1\kappa}\epsilon _{2\mu})(\tilde\epsilon _{1\beta}\tilde\epsilon _{2\sigma}- \tilde\epsilon _{1\sigma}\tilde\epsilon _{2\beta}) [ \rho^{(1)}_{12}({\bf q}+{\bf p})-  \rho^{(1)}_{21}({\bf q}+{\bf p})] \bigg\} .
\end{eqnarray*}
Given the above expressions it is easy to see that no mixing occurs between $\Delta _V$ and the set $\{\Delta _I,\Delta _Q,\Delta _U\}$ as a result of the damping term Eq.(\ref{eq:4.43}). Therefore, in our approximation, no circular polarization is generated by first order processes up to $\mathcal{O}(g^2)$. 

\section{The Second-Order Interaction Term}
\label{sec:V}

We now move on to calculate the contribution to the evolution of the photon density matrix from scattering processes which are second order in the interaction Hamiltonian density operator Eq.~(\ref{eq:4.34}). Details of the calculation of the second-order interaction Hamiltonian and the corresponding refractive term are presented in Appendix~\ref{sec-B}. The second-order interaction Hamiltonian is given by 
\begin{eqnarray}
	\label{eq:5.51}
		 \phantom{}^{2}\hat H_{\text{int}}(t)&=&-\frac{i(2\pi)^3}{2}(2g)^2\int \frac{d^3{\bf p}_1}{(2\pi )^3 2p_1^0}d{ l}_1 d{ l}_2\,\frac{d^3{\bf p}_2}{(2\pi )^3 2p_2^0} \epsilon^{\mu \nu \alpha \beta}\epsilon^{\rho \sigma \lambda \kappa }\tilde T_\nu( {l_1}) \tilde T_\sigma( {l_2}) (-ig_{\mu\rho}) \\
		\nonumber &\times &  \bigg[\big( ( p_{1}+{p_2}+{l_2})_\alpha (2p_2 +{l_2}) _\lambda \big) e^{-i((p_1)^0-({l_1})^0-({p_2})^0-({l_2})^0)t}\delta^3 ({\bf p}_1-{{\bf l}_1}-{\bf p}_2-{{\bf l}_2})\frac{\epsilon^* _{s\kappa}(p_2) \epsilon _{s\beta}(p_1)\hat a_s^\dagger (p_2)\hat a_r(p_1)}{(p_2+l_2)^2+i\epsilon}\\
\nonumber &&+\big( (p_{1}+ p_2 -l_2)_\alpha (2p_2 -l_2)_\lambda  \big)   e^{i((p_1)^0+{l_1}^0- (p_2)^0+{l_2}^0  )t}  \delta ^3({\bf p}_1+{\bf l}_1-{\bf p}_2+{\bf l}_2)\frac{\epsilon^* _{s\beta}(p_1) \epsilon _{r\kappa}(p_2)\hat a_s^\dagger (p_1)\hat a_r(p_2)}{(p_2-l_2)^2+i\epsilon}\bigg].
\end{eqnarray}
The refractive term of Eq.~(\ref{eq:3.27}) due to this interaction Hamiltonian is
\begin{eqnarray}
	\label{eq:5.52}
	\nonumber	  i\left \langle \left[ \phantom{}^{2}\hat H_{\text{int}}(0),\mathcal{\hat D}_{uv}({\bf q})\right] \right \rangle &=&i(2\pi)^3\delta^3(0) (2g)^2\big( \delta_{ur}\rho_{sv}({\bf q})-\delta_{vs}\rho_{ur}({\bf q})  \big)  \epsilon_s^\mu(q) \epsilon_r^\nu(q)  \mathcal{T}_{\mu \nu}(q),
\end{eqnarray}
where we have defined $\mathcal{T}_{i j}(q)$ via 
\begin{eqnarray}
	\label{eq:5.53}
		(2\pi)^3\delta^3(0) \mathcal{T}_{i j}(q)&=& -\int d{ l}_1 d{ l}_2\, (2\pi)^3 \delta^3 ({{\bf l}_1}+{{\bf l}_2})\tilde T_j(l_2^0,{\bf l}_2)\tilde T_i(l_1^0,{\bf l}_2)  [ q\cdot (l_1 +l_2)]\\
\nonumber &&\qquad\times\bigg[\frac{1}{(l_2)^2+2l_2\cdot q+i\epsilon}- \frac{1}{(l_2)^2-2l_2\cdot q+i\epsilon}\bigg] .
\end{eqnarray}
Note that the interaction Hamiltonian Eq.~(\ref{eq:5.51}) and the refractive term Eq.~(\ref{eq:5.52}) were computed using the photon propagator in the Feynman gauge Eq.~(\ref{eq:B-3}), but a demonstration of gauge invariance of these results is detailed in Appendix~\ref{sec-C}. For an unpolarized photon ensemble, it is easy to see that the right side of Eq.(\ref{eq:5.52}) vanishes when evaluated in terms of $\rho^{(0)}$ 
defined in Eq.~(\ref{eq:3.28}) above. The contribution to the perturbed density matrix, $\rho^{(1)}$, to second order in the interaction coupling $g$ is 
\begin{eqnarray} 
	\label{eq:5.54}
		 \frac{d} {dt} \rho^{(1)}_{uv}({\bf q}) &=&\frac{2ig^2}{q^0}\big( \delta_{ur}\rho^{(1)}_{sv}({\bf q})-\delta_{vs}\rho^{(1)}_{ur}({\bf q})  \big)  \epsilon_s^\mu(q) \epsilon_r^\nu(q)  \mathcal{T}_{\mu \nu}(q).
\end{eqnarray}
Here we have used Eq.~(\ref{eq:3.27}) and have ignored the damping term, which is of $\mathcal{O}(g^4)$. 
Expressing Eq.~(\ref{eq:5.54}) explicitly in terms of photon density matrix components gives the
evolution equations
\begin{eqnarray}
	\label{eq:5.55}
		 \frac{d} {dt} \rho^{(1)}_{11}({\bf q}) &=& \frac{2ig^2}{q^0}\big( -\epsilon_1^\mu\epsilon_2^\nu\rho^{(1)}_{12}({\bf q})+ \epsilon_2^\mu\epsilon_1^\nu\rho^{(1)}_{21}({\bf q})  \big) \mathcal{T}_{\mu \nu}(q)\\
	\label{eq:5.56}
		\frac{d} {dt} \rho^{(1)}_{22}({\bf q}) &=& \frac{2ig^2}{q^0}\big( \epsilon_1^\mu\epsilon_2^\nu \rho^{(1)}_{12}({\bf q})-\epsilon_2^\mu\epsilon_1^\nu\rho^{(1)}_{21}({\bf q})  \big) \mathcal{T}_{\mu \nu}(q) \\		 
	\label{eq:5.57}
		\frac{d} {dt} \rho^{(1)}_{12}({\bf q}) &=& \frac{2ig^2}{q^0}\big( [\epsilon_1^\mu \epsilon_1^\nu -\epsilon_2^\mu \epsilon_2^\nu ]\rho^{(1)}_{12}({\bf q})- \epsilon_2^\mu \epsilon_1^\nu[\rho^{(1)}_{11}({\bf q})-\rho^{(1)}_{22}({\bf q})] \big) \mathcal{T}_{\mu \nu}(q)\\
	\label{eq:5.58}
		 \frac{d} {dt} \rho^{(1)}_{21}({\bf q}) &=& \frac{2ig^2}{q^0}\big(  -[\epsilon_1^\mu \epsilon_1^\nu -\epsilon_2^\mu \epsilon_2^\nu ]\rho^{(1)}_{21}({\bf q})+\epsilon_1^\mu\epsilon_2^\nu[\rho^{(1)}_{11}({\bf q}) - \rho^{(1)}_{22}({\bf q})] \big) \mathcal{T}_{\mu \nu}(q).
\end{eqnarray}
All polarization vectors $\epsilon_r^\mu$ in the above expression depend on
${\bf q}$,  the same photon momentum as in the argument of the photon density matrix. Using the Stokes brightness perturbations in Eqs.~(\ref{eq:3.30})--(\ref{eq:3.33}), we can now express the evolution of the polarization of the photon ensemble due to processes mediated by $\phantom{}^{2}\hat H_{\text{int}}(t)$ as
\begin{eqnarray}
	\label{eq:5.59}
		\frac{d} {dt} \Delta _I({\bf q}) &=& 0,\\
	\label{eq:5.60}
		\frac{d} {dt} \Delta _Q({\bf q}) &=& -\frac{g^2}{q^0}\left( \zeta({\bf\hat q})\Delta _V({\bf q}) +i\psi({\bf\hat q})\Delta _U({\bf q})\right)  , \\
		\label{eq:5.61}
		\frac{d} {dt} \Delta _U({\bf q}) &=& -\frac{g^2}{q^0}\left( \chi({\bf\hat q})\Delta _V({\bf q}) -i\psi({\bf\hat q})\Delta _Q({\bf q}) \right), \\
		\label{eq:5.62}
		\frac{d} {dt} \Delta _V({\bf q}) &=& \frac{g^2}{q^0}
		\left(\chi({\bf\hat q})\Delta _U({\bf q})+\zeta({\bf\hat q})\Delta_Q({\bf q})\right), 
\end{eqnarray}
where we have defined the contractions
\begin{eqnarray}
\zeta({\bf\hat q})&\equiv& -2\mathcal{T}_{\mu \nu}(\epsilon _{1}^\mu\epsilon _2 ^\nu +\epsilon _{2}^\mu\epsilon _1^\nu) ,\\
\chi({\bf\hat q})&\equiv& 2\mathcal{T}_{\mu \nu}(\epsilon _1^{\mu}\epsilon _1 ^\nu -\epsilon _2^{\mu}\epsilon _2^\nu) \\
\psi({\bf\hat q})&\equiv& 2\mathcal{T}_{\mu \nu}(\epsilon _{1}^\mu\epsilon _2 ^\nu -\epsilon _{2}^\mu\epsilon _1^\nu )
\end{eqnarray}
and the quantity $\mathcal{T}_{\mu \nu}$ is defined via the integral expression Eq.~(\ref{eq:5.53}).

The circular polarization brightness $\Delta_V$ is sourced by terms which are proportional to either 
$\Delta_Q$ or $\Delta_U$: a linearly polarized photon ensemble in the presence of the interaction 
Eq.~(\ref{eq:4.34}) will acquire circular polarization due to processes which are of second order in the interaction. As long as the interaction acts, both Stokes brightnesses $\Delta_Q$ and $\Delta_U$
are rotated with $\Delta_V$. Note that the above equations do not depend on the time component of $T_\mu(x)$ since the polarization vectors are purely spatial.

\section{Discussion}
\label{sec:VI} 

The evolution equations (\ref{eq:5.59}) to (\ref{eq:5.62}), along with Eqs.~(\ref{eq:4.46}) to (\ref{eq:4.49}), 
are the central result of this paper. Other source terms associated with the usual Compton
scattering effects will also appear on the right sides. While the polarization brightnesses will
be zero prior to recombination, during recombination $\Delta_Q$ and $\Delta_U$ become
non-zero, with an amplitude a factor of 20 smaller than the intensity fluctuations $\Delta_I$. It is
easy to see by inspection of the evolution equations 
that at that point, $\Delta_Q$ and $\Delta_V$ will rotate into each other
with a characteristic angular frequency $\omega_{QV} = g^2\zeta/k^0$, $\Delta_U$ and
$\Delta_V$ will rotate into each other with a characteristic frequency $\omega_{UV} = g^2\chi/k^0$,
and $\Delta_Q$ and $\Delta_U$ will rotate into each other with a characteristic
frequency $\omega_{QU}=g\alpha$, along with an exponential decay or growth of
$\Delta_Q$ and $\Delta_U$ associated with the first order damping effects. All of
these source terms are active whenever the interaction Eq.~(\ref{eq:4.34}) is nonzero, in contrast
to the conventional Compton scattering terms, which are only significant when the
photons propagate through ionized regions of the universe. 

The rotation between $\Delta_Q$ and $\Delta_U$ can be constrained from
current measurements in a straightforward way. Linear polarization on the
sky is conveniently expressed in a different basis, corresponding to
the ``gradient'' and ``curl'' pieces of the polarization tensor field \cite{kam97a,kam97b},
also known as the E/B decomposition \cite{sel97}. This decomposition is useful
because scalar density perturbations in the universe, which evolve into the structures
we see today via gravitational instability, generate only E-mode polarization. Subsequent
rotation of the polarization plane as the wave propagates rotates E-mode into B-mode.
Current limits on the amplitude of B-mode polarization (see Refs.\ \cite{nol08,ade08,bis08}
for some recent linear polarization measurements) can be translated
into limits on the total rotation of linear polarization between the time of last scattering
and today; see Refs.~\cite{kos05,kahniashvili:2009:b} for corresponding limits on magnetic fields
due to Faraday rotation. Precise limits on the interaction studied here from the first-order
rotation effect Eqs.~(\ref{eq:4.47}) and (\ref{eq:4.48}) can be obtained similarly, and will
be computed elsewhere. But we know that the total amount of rotation must be
small at frequencies between 50 GHz and 150 GHz where good measurements
of the primordial linear polarization have been made.  
Given this observational constraint on linear polarization rotation, 
can some realistic cosmological model
generate detectable circular polarization via Eqs.~(\ref{eq:5.60})--(\ref{eq:5.62})?
First, note that $\omega_{QU}$ has dimensions of $[g T]$, and that for
this rotation to be below current limits, 
\begin{equation}
\omega_{QU} \simeq g T\ll H_0; 
\label{omegaQU_limit}
\end{equation}
otherwise, as the microwave background photons propagate a Hubble distance from
last scattering until today, we would have substantial rotation of E-mode into B-mode
polarization.  We are aware of no other laboratory or theoretical constraints
on this class of interactions. 

Now note that any first order damping effects will contribute an extra factor of length compared to $\alpha(q)$,
arising from an additional time integral over the field $T_\mu(x)$; if the field is active
for all times, this leads to roughly a factor of $H_0^{-1}$. Then
the time scale for exponential growth or decay of
linear polarization is, by dimensional estimate, $\omega_{QU}(\omega_{QU}/H_0)$,
which is small compared to $\omega_{QU}$: we can always neglect the exponential growth
or decay of linear polarization compared to its rotation. 

For generation of $V$ polarization, $\omega_{QV}$ and $\omega_{UV}$ both have
dimensions $[g^2 T^2/k^0]$, differing only by a geometric factor related to
the propagation directions of the photons and the direction of the field $T^\mu(x)$. 
So a dimensional estimate for both is $\omega_{QV}\simeq\omega_{UV}\simeq
\omega_{QU}(\omega_{QU}/k^0)$. A typical microwave background photon today
will have a frequency of $k^0\simeq 100\,{\rm GHz}$ or $10^{11}\,{\rm s}^{-1}$, while the characteristic
size of $\omega_{QU}$ at the observational limit is $H_0$, a huge mismatch in
scales. So in the case considered here, the generation of circular polarization
is always vastly subdominant to the rotation of the linear polarization, and can be
neglected. 

The calculation presented here demonstrates generally 
that, given additional interactions beyond Compton scattering,
circular polarization is not necessarily zero, and elaborates the 
framework for calculating it for a given microphysical interaction. Other
interactions may well induce circular polarization without optical activity
from the linear interaction term, and for these cases circular polarization could be the
most constraining probe. We have also only considered a constant field $\mathcal{T}^\mu$
for simplicity; calculations for a non-constant field are messier but straightforward, involving
convolutions over the field and photon distributions. 
Spatial or temporal variations in the field could change the relative importance of the
optical activity and circular polarization generation effects. 
In particular, the torsion field necessarily couples to fermions via the interaction $\mathcal{L}_{TF} = g_{1}\mathcal{T}_{\mu} \bar{\psi}\gamma^{\mu} \psi$ as well as a torsion-induced four-fermion interaction $\mathcal{L}_{FF} = g_{2} \bar{\psi}\gamma^{5}\gamma^{\mu} \psi \bar{\psi}\gamma_{5}\gamma_{\mu} \psi$ \cite{Shapiro,Perez:2005pm,sy}, where $g_{1}$ and $g_{1}$ are renormalized couplings.  It would be interesting to study the coupled system of torsion and fermions subject to our formalism for evaluating V, since backreaction effects from the torsion-fermion interaction could enhance the amplitude and modify the scale of V; we leave this to future work.

We encourage experimenters to make measurements testing
the standard lore that circular polarization of the cosmic microwave background
radiation should be identically zero, and theorists to consider the effects  
of any non-standard photon couplings on microwave background polarization as
photons propagate over cosmological distances.

\begin{acknowledgments}
S.A.\ is partly supported by a CAREER
grant from the National Science Foundation. J.O.\ is supported by the Eberly College of Science and acknowledges support from the Alfred P. Sloan Foundation.  A.K.\ has been partly supported by
NSF grant AST-0546035. 
\end{acknowledgments}
\appendix

\renewcommand{\theequation}{A-\arabic{equation}}
\setcounter{equation}{0}  

\section{First-Order Calculation}
\label{sec-A} 

In this appendix we determine the contribution to the time evolution of the photon density matrix due to processes which are first order in the interaction Hamiltonian density Eq.~(\ref{eq:4.34}). We first detail the calculation of the refractive term and then the damping term of Eq.~(\ref{eq:3.27}). The first-order interaction Hamiltonian is simply given by
\begin{equation}
	\label{eq:A-1}
		\hat H^{(1)}_{\text{int}}(t)=\int d^3  {\bf x} \phantom{1}\mathcal{\hat H}_{T} \,.
\end{equation}
In the approximation employed here, all operators in the collision terms of Eq.~(\ref{eq:3.27}) are from the free theory. From this point on, all operators represent free theory operators and we drop the `0' superscript used in Sec.\ref{sec:III}. Ignoring any processes in which two physical photons are either annihilated or created leads to the following expression for the interaction Hamiltonian:
\begin{eqnarray}
	\label{eq:A-2}
		\phantom{}^{1}\hat H_{\text{int}}(t) &=& 2g\int d^3 {\bf x} \, dp \,d {\bf k}\,  d{\bf k}' \,
		 \epsilon^{\mu\nu\alpha\beta} \phantom{1}\tilde T_\nu(p)(-ik'_\alpha )\nonumber\\
&&\qquad\qquad \times\left(\hat a_s^\dagger (k)\hat a_r (k') \epsilon^* _{s\mu}(k) \epsilon _{r\beta}(k')
e^{i(p+k-k')\cdot x}
 -\hat a_r^\dagger (k')\hat a_s(k)  \epsilon^* _{r\beta}(k')\epsilon_{s\mu}(k)e^{i(p+k'-k)\cdot x}\right) \,,
\end{eqnarray}
where we have used the expression for $\hat A_\mu$ given in Eq.~(\ref{eq:3.14}) and we have used the shorthand notation
\begin{equation}
		\nonumber \int dp  \equiv \int \frac{d^4p}{(2\pi )^4} \,,\qquad\qquad
		\int d {\bf k}  \equiv \int \frac{d^3 {\bf k}}{(2\pi )^3 2 k^0} \,.
\end{equation}
Performing the spatial integral and relabeling dummy momentum variables and spacetime indices in the second term of (\ref{eq:A-2}) gives
\begin{equation}
	\label{eq:A-3}
		\phantom{}^{1} \hat H_{\text{int}}(t) = -2ig\int dp\,d{\bf k}\,d{\bf k}' (2\pi )^3 \delta ^{(3)}({\bf k} +{\bf p} -{\bf k}')
  e^{i(k_0+p_0-k'_0)t} \epsilon^{\mu\nu\alpha\beta}\tilde T_\nu(p)\hat a_s^\dagger (k)\hat a_r (k') \epsilon^* _{s\mu}(k)(k+k')_\alpha \epsilon _{r\beta}(k') \,.
\end{equation}
Now perform the $\int d{\bf k}'$ integral, after which we have
\begin{eqnarray}
\label{eq:A-4}
\nonumber	\phantom{}^{1} \hat H_{\text{int}}(t) &=& -2 ig\int dp\,d{\bf k}\, \frac{\tilde T_\nu(p) e^{i(k_0+p_0-\left|{\bf k}+{\bf p}\right|)t}}{2\left|{\bf k}+{\bf p}\right|} \hat a_s^\dagger (k)\hat a_r (\tilde k) \epsilon^* _{s\mu}(k)\big(\epsilon^{\mu\nu 0\beta}(\left|{\bf k}\right|+\left|{\bf k}+{\bf p}\right|)+\epsilon^{\mu\nu j\beta}(2{k}_j+{p}_j) \big)\epsilon _{r\beta}(\tilde k ) \,,\\
	\label{eq:A-5}
		&=& -2 ig\int dp\,d{\bf k}\, \frac{\epsilon^{\mu\nu\alpha\beta}\tilde T_\nu(p) e^{i(k_0+p_0-\left|{\bf k}+{\bf p}\right|)t}}{2\left|{\bf k}+{\bf p}\right|} \hat a_s^\dagger (k)\hat a_r (\tilde k) \epsilon^* _{s\mu}(k)(k+\tilde k)_\alpha \epsilon _{r\beta}(\tilde k) \,.
\end{eqnarray}
where $(\tilde k)_\alpha =(\left|{\bf k}+{\bf p}\right|,{\bf k}+{\bf p})$. Equation~(\ref{eq:A-5}) is our first-order interaction Hamiltonian. Now the commutator necessary for the refractive term of Eq.~(\ref{eq:3.27}) is given by
\begin{eqnarray}
	\label{eq:A-6}
		\left[ \phantom{}^{1}\hat H_{\text{int}}(t),\mathcal{\hat D}_{uv}({\bf q})\right] &=& -2 ig\int dp\,d{\bf k}\, \frac{\tilde T_\nu(p) e^{i(k_0+p_0-\left|{\bf k}+{\bf p}\right|)t}}{2\left|{\bf k}+{\bf p}\right|}  \epsilon^* _{s\mu}(k)\big(\epsilon^{\mu\nu \alpha\beta}(k+\tilde k)_\alpha \big)\epsilon _{r\beta}(\tilde k )\\
\nonumber &&\times(2\pi )^32q^0\left( \delta_{ur}\delta^{3}({\bf q} - {\bf k}-{\bf p} )\hat a_s^\dagger (k)\hat a_v (q)-\delta_{vs}\delta^{3}({\bf q} - {\bf k} )\hat a_u^\dagger (q)\hat a_r (\tilde k) \right) \,
\end{eqnarray}
where we have used the canonical commutation relations between the free creation and annihilation operators Eq.~(\ref{eq:3.15}). Taking the expectation value of (\ref{eq:A-6} and using the relationship between the number operator and density matrix given by Eq.~(\ref{eq:3.17}), we arrive at the following expression:
\begin{eqnarray}
	\label{eq:A-7}
		i\left\langle\left[ \phantom{}^{1}\hat H_{\text{int}}(t),\mathcal{\hat D}_{uv}({\bf q})\right]\right\rangle &=&2g\int dp\,d{\bf k}\, \frac{\tilde T_\nu(p) e^{i(k_0+p_0-\left|{\bf k}+{\bf p}\right|)t}}{2\left|{\bf k}+{\bf p}\right|}  \epsilon^* _{s\mu}(k)\epsilon^{\mu\nu \alpha\beta}(k+\tilde k)_\alpha \epsilon _{r\beta}(\tilde k )\\
\nonumber &&\times(2\pi )^6(2q^0)^2\delta^{3}({\bf q} - {\bf k}-{\bf p} )\delta^{3}({\bf q} - {\bf k} )\left( \delta_{ur}\rho_{sv}( {\bf q})-\delta_{vs}\rho_{ur}({\bf q})\right) \,.
\end{eqnarray}
Perform the $\int d{\bf k}$ integral gives
\begin{eqnarray}
	\label{eq:A-8}
		i\left\langle\left[ \phantom{}^{1}\hat H_{\text{int}}(t),\mathcal{\hat D}_{uv}({\bf q})\right]\right\rangle &=&2g\int dp\, \frac{\tilde T_\nu(p) e^{i(q_0+p_0-\left|{\bf q}+{\bf p}\right|)t}}{4\left|{\bf q}\right|\left|{\bf q}+{\bf p}\right|}  \epsilon^* _{s\mu}(q)\epsilon^{\mu\nu \alpha\beta}(q+\tilde q)_\alpha \epsilon _{r\beta}(\tilde q )\\
\nonumber &&\times(2\pi )^3(2q^0)^2\delta^{3}(-{\bf p} )\left( \delta_{ur}\rho_{sv}( {\bf q})-\delta_{vs}\rho_{ur}({\bf q})\right) \,,
\end{eqnarray}
where as before we have defined $(\tilde q)_\alpha =(\left|{\bf q}+{\bf p}\right|,{\bf q}+{\bf p})$ so that the above can be expressed as 
\begin{eqnarray}
	\label{eq:A-9}
		i\left\langle\left[ \phantom{}^{1}\hat H_{\text{int}}(t),\mathcal{\hat D}_{uv}({\bf q})\right]\right\rangle &=&2g\epsilon^* _{s\mu}({\bf q})(2q^0)\left( \delta_{ur}\rho_{sv}( {\bf q})-\delta_{vs}\rho_{ur}({\bf q})\right)\int dp\, (2\pi )^3\delta^{3}(-{\bf p} )\frac{\tilde T_\nu(p) e^{i(q_0+p_0-\left|{\bf q}+{\bf p}\right|)t}}{2\left|{\bf q}+{\bf p}\right|} \\
\nonumber &&\times \big(\epsilon^{\mu\nu 0\beta}(\left|{\bf q}\right|+\left|{\bf q}+{\bf p}\right|)+\epsilon^{\mu\nu j\beta}(2{q}_j+{p}_j) \big)\epsilon _{r\beta}({\bf q}+ {\bf p}) \,.
\end{eqnarray}
Define the quantity
\begin{eqnarray}
	\label{eq:A-10}
\nonumber	(2\pi)^3\delta^3(0)\mathcal{ A}^{\mu \beta}(q) &=&(2\pi)^3\delta^3(0)\mathcal{ A}^{[\mu \beta]}(q)\,,\\
&= &\int dp\, (2\pi )^3\delta^{3}(-{\bf p} )\frac{\tilde T_\nu(p) }{2\left|{\bf q}+{\bf p}\right|} \big(\epsilon^{\mu\nu 0\beta}(\left|{\bf q}\right|+\left|{\bf q}+{\bf p}\right|)+\epsilon^{\mu\nu j\beta}(2{q}_j+{p}_j) \big) \,,
\end{eqnarray}
where we have anticipated the fact that the delta-distribution factor will be present once an appropriate form for the external field $T_\mu(x)$ is chosen. The refractive term of Eq.(\ref{eq:3.27}) 
due to first order processes can then be expressed as
\begin{eqnarray}
	\label{eq:A-11}
		i\left\langle\left[ \phantom{}^{1}\hat H_{\text{int}}(0),\mathcal{\hat D}_{uv}({\bf q})\right]\right\rangle &=&4g q^0 (2\pi)^3\delta^3(0) \epsilon^* _{s\mu}({\bf q})\left( \delta_{ur}\rho_{sv}( {\bf q})-\delta_{vs}\rho_{ur}({\bf q})\right)\mathcal{ A}^{\mu \beta}(q) \epsilon _{r\beta}({\bf q}) \,.
\end{eqnarray}

Next we consider the damping term of Eq.~(\ref{eq:3.26}), the integrand of which involves the double commutator
\begin{eqnarray}
	\label{eq:A-12}
	\nonumber &&\left[ \phantom{}^{1}\hat H_{\text{int}}(t-t'),\left[ \phantom{}^{1}\hat H_{\text{int}}(t),\mathcal{\hat D}_{uv}({\bf q})\right]\right]  =  (2ig)^2\int dl\,dp\,d{\bf k}_2\,d{\bf k}_1\, \frac{\tilde T_\nu(p) e^{i((k_1)_0+p_0-\left|{\bf k}_1+{\bf p}\right|)t}}{2\left|{\bf k}_1+{\bf p}\right|} \frac{\tilde T_\lambda(l) e^{i((k_2)_0+l_0-\left|{\bf k}_2+{\bf l}\right|)(t-t')}}{2\left|{\bf k}_2+{\bf l}\right|} \\ 
 &&\qquad\qquad\times   \epsilon^* _{s\mu}(k_1)\epsilon^{\mu\nu \alpha\beta}(k_1+\tilde k_1)_\alpha \epsilon _{r\beta}(\tilde k_1 ) \epsilon^* _{n\sigma}(k_2)\epsilon^{\sigma\lambda\rho\kappa}(k_2+\tilde k_2)_\rho \epsilon _{m\kappa}(\tilde k_2 )\\
\nonumber &&\qquad\qquad\qquad\times(2\pi )^62q^0\bigg( \delta_{ur}\delta^{3}({\bf q} - {\bf k}_1-{\bf p})\big( 2(k_1)^0\delta_{ms}\delta^{3}({\bf k}_2+{\bf l} - {\bf k}_1 ) \hat a_n^\dagger (k_2) \hat a_v (q)\\
\nonumber &&\qquad\qquad\qquad\qquad\qquad\qquad\qquad\qquad\qquad\qquad - 2q^0\delta_{nv}\delta^{3}({\bf k}_2 - {\bf q} )\hat a_s^\dagger (k_1)\hat a_m (\tilde k_2)\big)\\
\nonumber &&\qquad\qquad\qquad\qquad\qquad\qquad-\delta_{vs}\delta^{3}({\bf q} - {\bf k}_1 )\big( 2q^0\delta_{mu}\delta^{3}({\bf k}_2+{\bf l} - {\bf q} ) \hat a_n^\dagger (k_2) \hat a_r (\tilde k_1)\\
\nonumber &&\qquad\qquad\qquad\qquad\qquad\qquad\qquad\qquad\qquad\qquad - 2(k_2)^0\delta_{nr}\delta^{3}({\bf k}_2 - {\bf k}_1-{\bf p} )\hat a_u^\dagger (q)\hat a_m (\tilde k_2)\big) \bigg) \,,
\end{eqnarray}
where we have used the canonical commutation relations between the free creation and annihilation operators, Eq.~(\ref{eq:3.15}), and have defined  $(\tilde k_1)_\alpha =(\left|{\bf k}_1+{\bf p}\right|,{\bf k}_1+{\bf p})$ and $(\tilde k_2)_\alpha =(\left|{\bf k}_2+{\bf l}\right|,{\bf k}_2+{\bf l})$. Now the $\int d{\bf k}_1$ and the $\int d{\bf k}_2$ integrals can be performed, giving
\begin{eqnarray}
	\label{eq:A-13}
&&\left[ \phantom{}^{1}\hat H_{\text{int}}(t-t'),\left[ \phantom{}^{1}\hat H_{\text{int}}(t),\mathcal{\hat D}_{uv}({\bf q})\right]\right]  =  (2ig)^2 2q^0\int dl\,dp\, \tilde T_\lambda(l) \epsilon^{\sigma\lambda\rho\kappa} \tilde T_\nu(p) \epsilon^{\mu\nu \alpha\beta}\\ 
	\nonumber &&\qquad\times\bigg(\frac{ e^{i(\left|{\bf q}-{\bf p}\right|+p_0-\left|{\bf q}\right|)t}\epsilon^* _{s\mu}(\tilde q_{-p})(\tilde q_{-p}+ q)_\alpha \epsilon _{r\beta}(q )\delta_{ur}}{4\left|{\bf q}-{\bf p}\right|\left|{\bf q}\right| }\\
	\nonumber &&\qquad\qquad\times \bigg[ 2\left|{\bf q}-{\bf p}\right|\frac{e^{i(\left|{\bf q} -{\bf p}-{\bf l}\right|+l_0-\left|{\bf q} -{\bf p}\right|)(t-t')}}{4\left|{\bf q} -{\bf p}-{\bf l}\right|\left|{\bf q} -{\bf p}\right|} \epsilon^* _{n\sigma}({\tilde q}_{-p-l})({ \tilde q}_{-p-l}+{ \tilde q}_{-p})_\rho \epsilon _{m\kappa}({\tilde q}_{-p} )\delta_{ms} \hat a_n^\dagger ({\tilde q}_{-p-l}) \hat a_v (q)\\
	\nonumber &&\qquad\qquad\qquad\qquad- 2q^0\frac{e^{i({q}_0+l_0-\left|{\bf q}+{\bf l}\right| )(t-t')}}{4\left|{\bf q}\right|\left|{\bf q}+{\bf l}\right|} \epsilon^* _{n\sigma}(q)(q+\tilde q_{+l})_\rho \epsilon _{m\kappa}(\tilde q_{+l} )\delta_{nv}\hat a_s^\dagger (\tilde q_{-p})\hat a_m (\tilde q_{+l})\bigg ]\\
	\nonumber &&\qquad\qquad -\frac{e^{i((q)_0+p_0-\left|{\bf q}+{\bf p}\right|)t}\epsilon^* _{s\mu}(q)(q+\tilde q_{+p})_\alpha \epsilon _{r\beta}(\tilde q_{+p} )\delta_{vs}}{4\left|{\bf q}+{\bf p}\right|\left|{\bf q}\right|}\\
	\nonumber &&\qquad\qquad\qquad\times \bigg[ 2q^0\frac{e^{i(\left|{\bf q}-{\bf l}\right|+l_0-\left|{\bf q}\right|)(t-t')}}{4\left|{\bf q}-{\bf l}\right|\left|{\bf q}\right|} \epsilon^* _{n\sigma}({ \tilde q}_{-l})({ \tilde q}_{-l} + q )_\rho \epsilon _{m\kappa}(q)\delta_{mu}\hat a_n^\dagger ({ \tilde q}_{-l}) \hat a_r (\tilde q_{+p})\\
	\nonumber &&\qquad\qquad\qquad\qquad - 2\left|{\bf q} +{\bf p}\right|\frac{e^{i(\left|{\bf q} +{\bf p}\right|+l_0 -\left|{\bf q} +{\bf p}+{\bf l}\right|)(t-t')}}{4\left|{\bf q} +{\bf p}\right|\left|{\bf q} +{\bf p}+{\bf l}\right|} \epsilon^* _{n\sigma}({ \tilde q}_{+p})({ \tilde q}_{+p}+{ \tilde q}_{+p+l})_\rho \epsilon _{m\kappa}({ \tilde q}_{+p+l} )\delta_{nr}\hat a_u^\dagger (q)\hat a_m ({ \tilde q}_{+p+l})\bigg ] \bigg) \,.
\end{eqnarray}
After taking the expectation value, this becomes
\begin{eqnarray}
	\label{eq:A-14}
&&\left\langle\left[ \phantom{}^{1}\hat H_{\text{int}}(t-t'),\left[ \phantom{}^{1}\hat H_{\text{int}}(t),\mathcal{\hat D}_{uv}({\bf q})\right]\right]\right\rangle  =  (2ig)^2 \int dl\,dp\, \tilde T_\lambda(l) \epsilon^{\sigma\lambda\rho\kappa} \tilde T_\nu(p) \epsilon^{\mu\nu \alpha\beta}(2\pi )^3\delta^3({\bf l} +{\bf p})e^{i(p_0 +l_0)t}\\ 
	\nonumber &&\qquad\times\bigg(\frac{ e^{-i(\left|{\bf q} \right|+l_0-\left|{\bf q} +{\bf l}\right|)t'}\epsilon^* _{s\mu}(\tilde q_{+l})(\tilde q_{+l}+ q)_\alpha \epsilon _{r\beta}(q )\delta_{ur}}{2\left|{\bf q}+{\bf l}\right| }\\
	\nonumber &&\qquad\qquad\times \bigg[  \epsilon^* _{n\sigma}({ q})({  q}+{ \tilde q}_{+l})_\rho \epsilon _{m\kappa}({\tilde q}_{+l} )\delta_{ms} \rho_{nv}({\bf q}) -  \epsilon^* _{n\sigma}(q)(q+\tilde q_{+l})_\rho \epsilon _{m\kappa}(\tilde q_{+l} )\delta_{nv} \rho_{sm}({\bf q}+{\bf l})\bigg ]\\
	\nonumber &&\qquad\qquad -\frac{e^{-i(\left|{\bf q}+{\bf p}\right|+l_0-\left|{\bf q}\right|)t'}\epsilon^* _{s\mu}(q)(q+\tilde q_{+p})_\alpha \epsilon _{r\beta}(\tilde q_{+p} )\delta_{vs}}{2\left|{\bf q}+{\bf p}\right|}\\
	\nonumber &&\qquad\qquad\qquad\times \bigg[  \epsilon^* _{n\sigma}({ \tilde q}_{+p})({ \tilde q}_{+p} + q )_\rho \epsilon _{m\kappa}(q)\delta_{mu}\rho_{nr}({\bf q}+{\bf p}) - \epsilon^* _{n\sigma}({ \tilde q}_{+p})({ \tilde q}_{+p}+{ q})_\rho \epsilon _{m\kappa}({ q} )\delta_{nr}\rho_{um}({\bf q})\bigg ] \bigg) \,.
\end{eqnarray}
Interchanging $l$ and $p$ in the first set of terms gives
\begin{eqnarray}
	\label{eq:A-15}
&&\left\langle\left[ \phantom{}^{1}\hat H_{\text{int}}(t-t'),\left[ \phantom{}^{1}\hat H_{\text{int}}(t),\mathcal{\hat D}_{uv}({\bf q})\right]\right]\right\rangle  =  (2ig)^2 \int dl\,dp\, (2\pi )^3\delta^3({\bf l} +{\bf p})e^{i(p_0 +l_0)t}\\ 
	\nonumber &&\qquad\times\bigg(\frac{ e^{-i(\left|{\bf q} \right|+p_0-\left|{\bf q} +{\bf p}\right|)t'}p_\lambda\phi(p) \epsilon^{\sigma\lambda\rho\kappa} l_\nu\phi(l) \epsilon^{\mu\nu \alpha\beta}\epsilon^* _{s\mu}(\tilde q_{+p})(\tilde q_{+p}+ q)_\alpha \epsilon _{r\beta}(q )\delta_{ur}}{2\left|{\bf q}+{\bf p}\right| }\\
	\nonumber &&\qquad\qquad\times \bigg[  \epsilon^* _{n\sigma}({ q})({  q}+{ \tilde q}_{+p})_\rho \epsilon _{m\kappa}({\tilde q}_{+p} )\delta_{ms} \rho_{nv}({\bf q}) -  \epsilon^* _{n\sigma}(q)(q+\tilde q_{+p})_\rho \epsilon _{m\kappa}(\tilde q_{+p} )\delta_{nv} \rho_{sm}({\bf q}+{\bf p})\bigg ]\\
	\nonumber &&\qquad\qquad -\frac{e^{-i(\left|{\bf q}+{\bf p}\right|+l_0-\left|{\bf q}\right|)t'}\tilde T_\lambda(l) \epsilon^{\sigma\lambda\rho\kappa} \tilde T_\nu(p) \epsilon^{\mu\nu \alpha\beta}\epsilon^* _{s\mu}(q)(q+\tilde q_{+p})_\alpha \epsilon _{r\beta}(\tilde q_{+p} )\delta_{vs}}{2\left|{\bf q}+{\bf p}\right|}\\
	\nonumber &&\qquad\qquad\qquad\times \bigg[  \epsilon^* _{n\sigma}({ \tilde q}_{+p})({ \tilde q}_{+p} + q )_\rho \epsilon _{m\kappa}(q)\delta_{mu}\rho_{nr}({\bf q}+{\bf p}) - \epsilon^* _{n\sigma}({ \tilde q}_{+p})({ \tilde q}_{+p}+{ q})_\rho \epsilon _{m\kappa}({ q} )\delta_{nr}\rho_{um}({\bf q})\bigg ] \bigg) \,.
\end{eqnarray}
After relabeling some spacetime and polarization indices in first set of terms, the above becomes
\begin{eqnarray}
	\label{eq:A-16}
&&\left\langle\left[ \phantom{}^{1}\hat H_{\text{int}}(t-t'),\left[ \phantom{}^{1}\hat H_{\text{int}}(t),\mathcal{\hat D}_{uv}({\bf q})\right]\right]\right\rangle  =  (2ig)^2 \int dl\,dp\, (2\pi )^3\delta^3({\bf l} +{\bf p})e^{i(p_0 +l_0)t}\\ 
	\nonumber &&\qquad\times\frac{ \tilde T_\nu(p) \epsilon^{\sigma\lambda\rho\kappa} \tilde T_\lambda(l) \epsilon^{\mu\nu \alpha\beta}\epsilon^* _{n\sigma}(\tilde q_{+p})(\tilde q_{+p}+ q)_\rho \epsilon _{m\kappa}(q )\epsilon^* _{s\mu}({ q})({  q}+{ \tilde q}_{+p})_\alpha \epsilon _{r\beta}({\tilde q}_{+p} )}{2\left|{\bf q}+{\bf p}\right| }\\
	\nonumber &&\qquad\qquad\times\bigg\{ e^{-i(\left|{\bf q} \right|+p_0-\left|{\bf q} +{\bf p}\right|)t'}\delta_{um}\bigg[  \delta_{rn} \rho_{sv}({\bf q}) -  \delta_{sv} \rho_{nr}({\bf q}+{\bf p})\bigg ] \\
\nonumber &&\qquad\qquad\qquad -e^{-i(\left|{\bf q}+{\bf p}\right|+l_0-\left|{\bf q}\right|)t'}\delta_{vs}\bigg[  \delta_{mu}\rho_{nr}({\bf q}+{\bf p}) - \delta_{nr}\rho_{um}({\bf q})\bigg ] \bigg\} \,.
\end{eqnarray}
Now integrate over $\int_{-t}^t dt'$ ($t\rightarrow \infty$) and define $\Delta q= \left|{\bf q} +{\bf p}\right|-\left|{\bf q} \right|$ to arrive at
\begin{eqnarray}
	\label{eq:A-17}
\nonumber D_{uv}  &\equiv &\int_{-\infty}^\infty dt'\left\langle\left[ \phantom{}^{1}\hat H_{\text{int}}(t'),\left[ \phantom{}^{1}\hat H_{\text{int}}(0),\mathcal{\hat D}_{uv}({\bf q})\right]\right]\right\rangle \\
&=&(2ig)^2  \int dl\,dp\, \frac{(2\pi )^4\delta^3({\bf l} +{\bf p}) }{2\left|{\bf q}+{\bf p}\right| }\tilde T_\nu(p) \epsilon^{\sigma\lambda\rho\kappa} \tilde T_\lambda(l) \epsilon^{\mu\nu \alpha\beta}\tilde\epsilon _{n\sigma}(\tilde q_{+p}+ q)_\rho ({  q}+{ \tilde q}_{+p})_\alpha \tilde\epsilon _{r\beta}\\
	\nonumber  && \qquad\qquad\qquad \times\bigg\{ \delta(p_0-\Delta q)\bigg[ \delta_{rn} \delta_{um} \epsilon _{m\kappa}\epsilon _{s\mu} \rho_{sv}({\bf q})-\delta_{um} \delta_{sv} \epsilon _{m\kappa}\epsilon _{s\mu}\rho_{nr}({\bf q}+{\bf p})\bigg]\\
\nonumber &&\qquad\qquad\qquad\qquad +\delta(l_0+\Delta q)\bigg[ \delta_{rn}\delta_{vs}\epsilon _{m\kappa}\epsilon _{s\mu}\rho_{um}({\bf q})-\delta_{um} \delta_{sv} \epsilon _{m\kappa}\epsilon _{s\mu}\rho_{nr}({\bf q}+{\bf p})\bigg]  \bigg\} \,,
\end{eqnarray}
where for convenience we have defined $D_{uv}$ above, as well as the abbreviations $\epsilon_{r\mu}=\epsilon_{r\mu}({\bf q})$ and $\tilde\epsilon_{r\mu}=\epsilon_{r\mu}({\bf q}+{\bf p})$. Note that since $\rho$ is expressed in a linear polarization basis all polarization vectors above have been assumed to be real.Now in the $\int d{\bf l} $ integral above, the relevant factor can be simplified as 
\begin{eqnarray*}
&&\tilde\epsilon _{n\sigma} \epsilon _{m\kappa}\epsilon _{s\mu} \tilde\epsilon _{r\beta}\int d{\bf l}\delta({\bf l}+ {\bf p})\tilde T_\lambda(l)({q}+{ \tilde q}_{+p})_\rho\epsilon^{\sigma\lambda\rho\kappa}\tilde T_\nu(p)({  q}+{ \tilde q}_{+p})_\alpha \epsilon^{\mu\nu \alpha\beta}\\
&&=\tilde\epsilon _{n\sigma} \epsilon _{m\kappa}\epsilon _{s\mu} \tilde\epsilon _{r\beta}\bigg\{ \tilde T_i(l_0,-{\bf p})[  \left |{\bf q}\right| + \left |{\bf q}+{\bf p}\right|] \epsilon^{\sigma i 0 \kappa} -\tilde T_0(l_0,-{\bf p})( 2{  q}_i +p_i) \epsilon^{\sigma i 0 \kappa}\bigg\}\\
&&\qquad \qquad\qquad\qquad \times\bigg\{ \tilde T_j(p_0,{\bf p})[  \left |{\bf q}\right| + \left |{\bf q}+{\bf p}\right| - p_0 ] \epsilon^{\mu j 0 \beta} - 2\tilde T_0(p_0,{\bf p}){  q}_j\epsilon^{\mu  j 0\beta} \bigg \} \, ,\\
&&=\tilde\epsilon _{n\sigma} \epsilon _{m\kappa}\epsilon _{s\mu} \tilde\epsilon _{r\beta}\bigg\{ \tilde T_i(l_0,{\bf p})[ ( \left |{\bf q}\right| + \left |{\bf q}+{\bf p}\right|)+ l_0] \epsilon^{\sigma i 0 \kappa} + 2\tilde T_0(l_0,{\bf p}){  q}_i \epsilon^{\sigma i 0 \kappa} \bigg\}\\
&&\qquad \qquad\qquad\qquad \times\bigg\{ \tilde T_j(p_0,{\bf p})[  \left |{\bf q}\right| + \left |{\bf q}+{\bf p}\right| - p_0 ] \epsilon^{\mu j 0 \beta} - 2\tilde T_0(p_0,{\bf p}){  q}_j\epsilon^{\mu  j 0\beta} \bigg \} \,,
\end{eqnarray*}
where in getting to the final expression we have used $p_{[\mu} \tilde T_{\nu]}(p)=0$, the fact that $\epsilon_{n\mu}(k)$ can be fixed as purely spatial for any momentum $k$, and the assumed behavior of $T_\mu(x)$ under a parity transformation, Eq.~(\ref{eq:4.36}). Now perform the appropriate energy integrals made trivial by the presence of delta-distributions in Eq.~(\ref{eq:A-17}). The relevant factor from the second term of Eq.~(\ref{eq:A-17}) becomes
\begin{eqnarray}
\label{eq:A-18}
\nonumber &&\tilde\epsilon _{n\sigma} \epsilon _{m\kappa}\epsilon _{s\mu} \tilde\epsilon _{r\beta}\int dl_0\delta(l_0+\Delta q) \bigg\{ \tilde T_i(l_0,{\bf p})[  \left |{\bf q}\right| + \left |{\bf q}+{\bf p}\right|+ l_0] \epsilon^{\sigma i 0 \kappa} + 2\tilde T_0(l_0,{\bf p}){  q}_i \epsilon^{\sigma i 0 \kappa} \bigg\}\\
\nonumber &&\qquad \qquad\qquad\qquad \qquad\qquad\qquad \times\bigg\{ \tilde T_j(p_0,{\bf p})[  \left |{\bf q}\right| + \left |{\bf q}+{\bf p}\right| - p_0 ] \epsilon^{\mu j 0 \beta} - 2\tilde T_0(p_0,{\bf p}){  q}_j\epsilon^{\mu  j 0\beta} \bigg \} \, ,\\
\nonumber &&=\tilde\epsilon _{n\sigma} \epsilon _{m\kappa}\epsilon _{s\mu} \tilde\epsilon _{r\beta} \bigg\{ \tilde T_i(-\Delta q,{\bf p})[  \left |{\bf q}\right| + \left |{\bf q}+{\bf p}\right|-\Delta q] \epsilon^{\sigma i 0 \kappa} + 2\tilde T_0(-\Delta q,{\bf p}){  q}_i \epsilon^{\sigma i 0 \kappa} \bigg\}\\
\nonumber &&\qquad \qquad\qquad\qquad \times\bigg\{ \tilde T_j(p_0,{\bf p})[  \left |{\bf q}\right| + \left |{\bf q}+{\bf p}\right| - p_0 ] \epsilon^{\mu j 0 \beta} - 2\tilde T_0(p_0,{\bf p}){  q}_j\epsilon^{\mu  j 0\beta} \bigg \} \, ,\\
\nonumber &&=\tilde\epsilon _{n\sigma} \epsilon _{m\kappa}\epsilon _{s\mu} \tilde\epsilon _{r\beta} \bigg\{ -2\tilde T_i(\Delta q,{\bf p})  \left |{\bf q}\right|  \epsilon^{\sigma i 0 \kappa} + 2\tilde T_0(\Delta q,{\bf p}){  q}_i \epsilon^{\sigma i 0 \kappa} \bigg\}\\
\nonumber &&\qquad \qquad\qquad\qquad \times\bigg\{ \tilde T_j(p_0,{\bf p})[  \left |{\bf q}\right| + \left |{\bf q}+{\bf p}\right| - p_0 ] \epsilon^{\mu j 0 \beta} - 2\tilde T_0(p_0,{\bf p}){  q}_j\epsilon^{\mu  j 0\beta} \bigg \} \, ,\\
&&=\tilde\epsilon _{n\sigma} \epsilon _{m\kappa}\epsilon _{s\mu} \tilde\epsilon _{r\beta} \bigg\{ -2\tilde T_\lambda(\Delta q,{\bf p}) q_\rho  \epsilon^{\sigma \lambda \rho \kappa}\bigg\}\bigg\{ \tilde T_j(p_0,{\bf p})[  \left |{\bf q}\right| + \left |{\bf q}+{\bf p}\right| - p_0 ] \epsilon^{\mu j 0 \beta} - 2\tilde T_0(p_0,{\bf p}){  q}_j\epsilon^{\mu  j 0\beta} \bigg \} \,,
\end{eqnarray}
where in arriving at the final expression above we have used the assumed behavior of $T_\mu(x)$ under a time reversal transformation Eq.~(\ref{eq:4.37}). Now perform the $p_0$ integral present in the first term of Eq. (\ref{eq:A-17}) and arrive at
\begin{eqnarray*}
&&\tilde\epsilon _{n\sigma} \epsilon _{m\kappa}\epsilon _{s\mu} \tilde\epsilon _{r\beta}\int dp_0\delta(p_0-\Delta q) \bigg\{ \tilde T_i(l_0,{\bf p})[  \left |{\bf q}\right| + \left |{\bf q}+{\bf p}\right|+ l_0] \epsilon^{\sigma i 0 \kappa} + 2\tilde T_0(l_0,{\bf p}){  q}_i \epsilon^{\sigma i 0 \kappa} \bigg\}\\
&&\qquad \qquad\qquad\qquad \qquad\qquad\qquad \times\bigg\{ \tilde T_j(p_0,{\bf p})[  \left |{\bf q}\right| + \left |{\bf q}+{\bf p}\right| - p_0 ] \epsilon^{\mu j 0 \beta} - 2\tilde T_0(p_0,{\bf p}){  q}_j\epsilon^{\mu  j 0\beta} \bigg \}  \\
&&=\tilde\epsilon _{n\sigma} \epsilon _{m\kappa}\epsilon _{s\mu} \tilde\epsilon _{r\beta} \bigg\{ \tilde T_i(l_0,{\bf p})[  \left |{\bf q}\right| + \left |{\bf q}+{\bf p}\right|+ l_0] \epsilon^{\sigma i 0 \kappa} + 2\tilde T_0(l_0,{\bf p}){  q}_i \epsilon^{\sigma i 0 \kappa} \bigg\}\bigg\{ 2\tilde T_\nu(\Delta q,{\bf p})   q_\alpha  \epsilon^{\mu \nu \alpha \beta}  \bigg \} \,.\\
\end{eqnarray*}
Next change the integration variable $l_0\rightarrow -p_0$ remaining in the first term of Eq.~(\ref{eq:A-17}) and the factor above becomes
\begin{eqnarray}
\label{eq:A-19}
&&\tilde\epsilon _{n\sigma} \epsilon _{m\kappa}\epsilon _{s\mu} \tilde\epsilon _{r\beta} \bigg\{ -\tilde T_i(p_0,{\bf p})[  \left |{\bf q}\right| + \left |{\bf q}+{\bf p}\right|- p_0] \epsilon^{\sigma i 0 \kappa} + 2\tilde T_0(p_0,{\bf p}){  q}_i \epsilon^{\sigma i 0 \kappa} \bigg\}\bigg\{ 2\tilde T_\nu(\Delta q,{\bf p})   q_\alpha  \epsilon^{\mu \nu \alpha \beta} \bigg \} \, ,
\end{eqnarray}
where we have again used Eq.~(\ref{eq:4.37}) . Following some relabeling of spacetime and polarization indices, the factor (\ref{eq:A-19}) becomes identical to the factor (\ref{eq:A-18}). Thus 
Eq.~(\ref{eq:A-17}) can be expressed as 
\begin{eqnarray}
	\label{eq:A-20}
&&D_{uv}  =  -(2ig)^2 \epsilon _{m\kappa}\epsilon _{s\mu} \int  \frac{dp }{2\left|{\bf q}+{\bf p}\right| } \tilde T_\lambda(\tilde p)q_\rho\epsilon^{\sigma\lambda\rho\kappa} \tilde T_\nu(p) (2{  q}+{ \tilde p})_\alpha \epsilon^{\mu\nu \alpha\beta}\tilde\epsilon _{n\sigma} \tilde\epsilon _{r\beta} \\
	\nonumber &&\qquad\qquad\qquad\times\bigg(  \delta_{nr} \delta_{us}  \rho_{mv}({\bf q}) + \delta_{rn}\delta_{vs}\rho_{um}({\bf q})-\delta_{us} \delta_{mv} \rho_{rn}({\bf q}+{\bf p})-\delta_{um} \delta_{sv} \rho_{nr}({\bf q}+{\bf p}) \bigg) \, ,
\end{eqnarray}
where we have defined the $4-$vector $\tilde p = (\Delta q, {\bf p})$ and recall that $\Delta q =\left|{\bf q}+{\bf p}\right|-\left|{\bf q}\right|$. The linear combinations necessary to describe the evolution of the independent polarization degrees of freedom are
\begin{eqnarray}
\label{eq:A-21}
D_{11}\pm D_{22}&=& -(2ig)^2  \int  \frac{dp }{2\left|{\bf q}+{\bf p}\right| } \tilde T_\lambda(\tilde p)q_\rho\epsilon^{\sigma\lambda\rho\kappa} \tilde T_\nu(p) (2{  q}+{ \tilde p})_\alpha \\
\nonumber &&\times\bigg\{(\tilde\epsilon _{1\sigma} \tilde\epsilon _{1\beta}+\tilde\epsilon _{2\sigma} \tilde\epsilon _{2\beta}) \bigg[ 2\epsilon _{1\kappa}\epsilon _{1\mu}  \rho_{11}({\bf q})\pm 2\epsilon _{2\kappa}\epsilon _{2\mu}  \rho_{22}({\bf q})\\
\nonumber &&\qquad\qquad\qquad\qquad\qquad+ (\epsilon _{2\kappa}\epsilon _{1\mu}\pm \epsilon _{1\kappa}\epsilon _{2\mu}) [ \rho_{12}({\bf q})+ \rho_{21}({\bf q})] \bigg]\\
\nonumber &&\qquad -\big(\epsilon _{1\kappa}\epsilon _{1\mu}\pm \epsilon _{2\kappa}\epsilon _{2\mu}\big)\bigg [ 2\tilde\epsilon _{1\beta}\tilde\epsilon _{1\sigma}\rho_{11}({\bf q}+{\bf p})+2\tilde\epsilon _{2\beta}\tilde\epsilon _{2\sigma}\rho_{22}({\bf q}+{\bf p})\\
\nonumber &&\qquad\qquad\qquad\qquad\qquad+(\tilde\epsilon _{1\beta}\tilde\epsilon _{2\sigma}+\tilde\epsilon _{2\beta}\tilde\epsilon _{1\sigma})[\rho_{12}({\bf q}+{\bf p})+\rho_{21}({\bf q}+{\bf p})]\bigg ] \bigg\} \,,\\
\label{eq:A-22}
D_{12}\pm D_{21}&=& -(2ig)^2  \int  \frac{dp }{2\left|{\bf q}+{\bf p}\right| } \tilde T_\lambda(\tilde p)q_\rho\epsilon^{\sigma\lambda\rho\kappa} \tilde T_\nu(p) (2{  q}+{ \tilde p})_\alpha \\
\nonumber &&\times\bigg\{(\tilde\epsilon _{1\sigma} \tilde\epsilon _{1\beta}+\tilde\epsilon _{2\sigma} \tilde\epsilon _{2\beta})\bigg[\epsilon _{1\kappa} \epsilon _{2\mu}[\rho_{11}({\bf q})\pm  \rho_{11}({\bf q})]+ \epsilon _{2\kappa}\epsilon _{1\mu} [ \rho_{22}({\bf q})\pm\rho_{22}({\bf q})]
\\
\nonumber &&\qquad\qquad\qquad\qquad\qquad +(\epsilon _{1\kappa} \epsilon _{1\mu}+\epsilon _{2\kappa}\epsilon _{2\mu}) [ \rho_{12}({\bf q})\pm\rho_{21}({\bf q})] \bigg]\\
\nonumber &&\qquad -(\epsilon _{2\kappa}\epsilon _{1\mu}\pm \epsilon _{1\kappa}\epsilon _{2\mu})\bigg[ \tilde\epsilon _{1\beta}\tilde\epsilon _{1\sigma}[\rho_{11}({\bf q}+{\bf p})\pm\rho_{11}({\bf q}+{\bf p})]+\tilde\epsilon _{2\beta}\tilde\epsilon _{2\sigma}[\rho_{22}({\bf q}+{\bf p}) \pm \rho_{22}({\bf q}+{\bf p})]\\
\nonumber &&\qquad\qquad\qquad\qquad\qquad\qquad +(\tilde\epsilon _{1\beta}\tilde\epsilon _{2\sigma}\pm \tilde\epsilon _{1\sigma}\tilde\epsilon _{2\beta}) [ \rho_{12}({\bf q}+{\bf p})\pm  \rho_{21}({\bf q}+{\bf p})] \bigg ] \bigg\} \, .
\end{eqnarray}
If we define the quantity 
\begin{eqnarray}
\label{eq:A-23}
(2\pi)^3\delta^{(3)}(0)\mathcal{B}_1^{\mu\kappa}(q)\equiv\int  \frac{dp }{2\left|{\bf q}+{\bf p}\right| } \tilde T_\lambda(\tilde p)q_\rho\epsilon^{\sigma\lambda\rho\kappa} \tilde T_\nu(p) (2{  q}+{ \tilde p})_\alpha \epsilon^{\mu\nu \alpha\beta}\tilde\epsilon _{n\sigma} \delta_{nr}\tilde\epsilon _{r\beta} \, 
\end{eqnarray} 
and the integral operator
\begin{eqnarray}
\label{eq:A-24}
(2\pi)^3 \delta^{(3)}(0)\mathcal{B}_2[q;\rho_{uv}]&\equiv& \epsilon _{m\kappa}\epsilon _{s\mu}\int  \frac{dp }{2\left|{\bf q}+{\bf p}\right| } \tilde T_\lambda(\tilde p)q_\rho\epsilon^{\sigma\lambda\rho\kappa} \tilde T_\nu(p) (2{  q}+{ \tilde p})_\alpha \epsilon^{\mu\nu \alpha\beta}\tilde\epsilon _{n\sigma} \tilde\epsilon _{r\beta}  \\
\nonumber &&\qquad\qquad\qquad\times\big(\delta_{us} \delta_{mv} \rho_{rn}({\bf q}+{\bf p})+\delta_{um} \delta_{sv} \rho_{nr}({\bf q}+{\bf p})\big) \, ,
\end{eqnarray} 
Eq.~(\ref{eq:A-20}) can be expressed in the general form
\begin{eqnarray}
\label{eq:A-25}
&& D_{uv}  =  -(2ig)^2(2\pi)^3\delta^{(3)}(0)\bigg\{ \epsilon_{m\kappa}\left(\epsilon_{u\mu}\rho_{m\nu}({\bf q}) + \epsilon_{v\mu}\rho_{um}({\bf q})\right) \mathcal{B}_1^{\mu\kappa}(q) -\mathcal{B}_2[q;\rho_{uv}(q)] \bigg\} \,.
\end{eqnarray}

\renewcommand{\theequation}{B-\arabic{equation}}
  \setcounter{equation}{0}  
\section{Second-Order Calculation}
\label{sec-B}

In this Appendix we explicitly calculate the relevant quantities describing the evolution of the photon density matrix due to processes which are second order in the interaction Hamiltonian Eq.~(\ref{eq:4.34}). The second-order scattering matrix operator is 
\begin{eqnarray}
		\nonumber \hat S^{(2)}&=&-\half\int_{-\infty}^\infty	 dt \int_{-\infty}^\infty	 dt' T \{ \hat H^{(1)}_{\text{int}}(t) \hat H^{(1)}_{\text{int}}(t')\} \,, \\
	\label{eq:B-1}
		&\equiv &-i\int_{-\infty}^\infty	 dt \hat H^{(2)}_{\text{int}}(t) \,,
\end{eqnarray}
where $\hat H^{(1)}_{\text{int}}(t)$ is the first-order interaction Hamiltonian, Eq.~(\ref{eq:A-1}). We will denote the interaction Hamiltonian operator which has a non-zero overlap with a single photon lying in both the initial and final scattering states as $\phantom{}^{2}H_{\text{int}}(t)$. Applying Wick's theorem to simplify the time ordered product in Eq.~(\ref{eq:B-1}) gives (ignoring vacuum terms)
\begin{eqnarray}
	\label{eq:B-2}
&&-i\hat H^{(2)}(t)=-\frac{1}{2}(2g)^2\int d^3{\bf x} d^4y\,\epsilon^{\mu \nu \alpha \beta}\epsilon^{\rho \sigma \lambda \kappa }T_{\nu}(x)T_{\sigma}(y):\bigg(
\contraction{}{\hat A}{_{\mu}(x)\partial_{\alpha}\hat A_{\beta}(x)}{\hat A}
\hat A_{\mu}(x)\partial_{\alpha}\hat A_{\beta}(x)\hat A_{\rho}(y)\partial_{\lambda}\hat A_{\kappa}(y) 
\nonumber \\
 &&\qquad + \contraction{}{\hat A}{_{\mu}(x)\partial_{\alpha}\hat A_{\beta}(x)\hat A_{\rho}(y)
 \partial_{\lambda}}{\hat A}
\hat A_{\mu}(x)\partial_{\alpha}\hat A_{\beta}(x)\hat A_{\rho}(y)\partial_{\lambda}\hat A_{\kappa}(y)
+\contraction{\hat A_{\mu}(x)\partial_{\alpha}}{\hat A}{_{\beta}(x)}{\hat A}
\hat A_{\mu}(x)\partial_{\alpha}\hat A_{\beta}(x)\hat A_{\rho}(y)\partial_{\lambda}\hat A_{\kappa}(y) 
+\contraction{\hat A_{\mu}(x)\partial_{\alpha}}{\hat A}{_{\beta}(x)\hat A_{\rho}(y)\partial_{\lambda}}{\hat A}\hat A_{\mu}(x)\partial_{\alpha}\hat A_{\beta}(x)\hat A_{\rho}(y)
\partial_{\lambda}\hat A_{\kappa}(y)\bigg): \, ,
\end{eqnarray}
where all partial derivatives are understood as acting solely on the function immediately to the right; the variable being differentiated is in the argument of this function. We denote the contraction of two operators $\hat A$ and $\hat B$ by
\begin{equation}
		\nonumber \contraction{}{A}{}{C}
				   \hat A \hat B \,.
\end{equation}
To simplify Eq.~(\ref{eq:B-2}), we use the free-theory photon propagator in the Feynman gauge, 
\begin{equation}
	\label{eq:B-3}
		D_{\mu\nu}(x-y)= \int\frac{d^4k}{(2\pi)^4}\frac{-ig_{\mu\nu}e^{-ik\cdot (x-y)}}{k^2+i\epsilon} \,.
\end{equation}
(See Appendix (\ref{sec-C}) below for a demonstration of gauge invariance, 
where we explicitly consider a different gauge-fixed photon propagator). In order to deal with the derivative couplings, we interpret the time ordering as $T^*$ ordering; specifically, we require that derivative couplings act outside of the time ordering operation \cite{sterman:1993:iqft}. For convenience, define the operators 
\begin{eqnarray}
	\label{eq:B-4}
		\hat A^+_{\mu}(x,p)&=& \hat a_s(p) \epsilon _{s\mu}(p)e^{-ip\cdot x} \,,\\
	\label{eq:B-5}	
		\hat A^-_{\mu}(x,p)&=& \hat a_s^\dagger (p) \epsilon^* _{s\mu}(p)e^{ip\cdot x} \,.
\end{eqnarray}
Equation (\ref{eq:B-2}) then becomes
\begin{eqnarray}
	\label{eq:B-6}
		\nonumber -i\hat H^{(2)}(t)&=&-\frac{1}{2}(2g)^2\int d^3{\bf x} d^4y \int \frac{d^4k}{(2\pi)^4} \frac{d^3{\bf p}_1}{(2\pi )^3 2p_1^0}\frac{d^3{\bf p}_2}{(2\pi )^3 2p_2^0} \epsilon^{\mu \nu \alpha \beta}\epsilon^{\rho \sigma \lambda \kappa }T_{\nu}(x)T_{\sigma}(y)\left( \frac{-ie^{-ik\cdot (x-y)}}{k^2+i\epsilon} \right) \\
		\nonumber &\times & :\bigg[ \phantom{1} g_{\mu\rho}(-ip_{1\alpha})\big( A^+_{\beta}(x,p_{1})-A^-_{\beta}(x,p_{1})\big)(-ip_{2\lambda})\big(A^+_{\kappa}(y,p_{2})-A^-_{\kappa}(y,p_{2})\big) \\
		&&\phantom{1}+(ik_{\lambda})g_{\mu\kappa}(-ip_{1\alpha})\big( A^+_{\beta}(x,p_{1})-A^-_{\beta}(x,p_{1})\big)\big( A^+_{\rho}(y,p_{2})+A^-_{\rho}(y,p_{2})\big)  \\
		\nonumber &&\phantom{1}+(-ik_{\alpha})g_{\beta\rho}\big( A^+_{\mu}(x,p_{1})+A^-_{\mu}(x,p_{1})\big)(-ip_{2\lambda})\big( A^+_{\kappa}(y,p_{2})-A^-_{\kappa}(y,p_{2})\big)  \\
		\nonumber &&\phantom{1}+(-ik_{\alpha})(ik_{\lambda})g_{\beta\kappa}\big( A^+_{\mu}(x,p_{1})+A^-_{\mu}(x,p_{1})\big)\big( A^+_{\rho}(y,p_{2})+A^-_{\rho}(y,p_{2})\big) \phantom{1}\bigg] : \, .
\end{eqnarray}
Picking out the non-vanishing overlap of $\hat H^{(2)}(t)$ on single-photon initial and final scattering states and calling this $\phantom{}^{2}\hat H_{\text{int}}(t)$ gives
\begin{eqnarray}
	\label{eq:B-7}
		\nonumber \phantom{}^{2}\hat H_{\text{int}}(t)&=&-\frac{i}{2}(2g)^2\int d^3{\bf x} d^4y \int dk\, d{\bf p}_1\,d{\bf p}_2 \epsilon^{\mu \nu \alpha \beta}\epsilon^{\rho \sigma \lambda \kappa }T_{\nu}(x)T_{\sigma}(y)\left( \frac{-ie^{-ik\cdot (x-y)}}{k^2+i\epsilon} \right) \\
		\nonumber &\times &  \bigg[ \phantom{1} g_{\mu\rho}(-ip_{1\alpha})(-ip_{2\lambda})\big( -\hat A^-_{\kappa}(y,p_{2}) \hat A^+_{\beta}(x,p_{1})-\hat A^-_{\beta}(x,p_{1})\hat A^+_{\kappa}(y,p_{2})\big) \\
		&&\phantom{1}+g_{\mu\kappa}(ik_{\lambda})(-ip_{1\alpha})\big( \hat A^-_{\rho}(y,p_{2})\hat A^+_{\beta}(x,p_{1})-\hat A^-_{\beta}(x,p_{1})\hat A^+_{\rho}(y,p_{2})\big)  \\
		\nonumber &&\phantom{1}+g_{\beta\rho}(-ik_{\alpha})(-ip_{2\lambda})\big( -\hat A^-_{\kappa}(y,p_{2}) \hat A^+_{\mu}(x,p_{1})+\hat A^-_{\mu}(x,p_{1})\hat A^+_{\kappa}(y,p_{2})\big)  \\
		\nonumber &&\phantom{1}+g_{\beta\kappa}(-ik_{\alpha})(ik_{\lambda})\big( \hat A^-_{\rho}(y,p_{2})\hat A^+_{\mu}(x,p_{1})+\hat A^-_{\mu}(x,p_{1})\hat A^+_{\rho}(y,p_{2})\big) \phantom{1}\bigg] \, ,
\end{eqnarray}
where we have again used the shorthand notation for the integral measures defined in Appendix (\ref{sec-A}). After some relabeling of spacetime indices, Eq.~(\ref{eq:B-7}) becomes
\begin{eqnarray}
	\label{eq:B-8}
		\nonumber \phantom{}^{2}\hat H_{\text{int}}(t)&=&-\frac{i}{2}(2g)^2\int d^3{\bf x} d^4y \int dk\, d{\bf p}_1\,d{\bf p}_2\, \epsilon^{\mu \nu \alpha \beta}\epsilon^{\rho \sigma \lambda \kappa }T_{\nu}(x)T_{\sigma}(y) \\
		&\times &  (-ig_{\mu\rho})\bigg[ \phantom{1}\left( \frac{e^{-ik\cdot (x-y)}}{k^2+i\epsilon} \right) \big( p_{1\alpha}(p_{2}-k)_{\lambda}-k_\alpha(p_2-k)_\lambda \big) \hat A^-_{\kappa}(y,p_{2}) \hat A^+_{\beta}(x,p_{1}) \\
		\nonumber &&\phantom{1}\phantom{1}\phantom{1}+\left( \frac{e^{-ik\cdot (y-x)}}{k^2+i\epsilon} \right)\big( p_{1\alpha}(p_2-k)_\lambda - k_\alpha (p_2-k)_\lambda \big) \hat A^-_{\beta}(x,p_{1})\hat A^+_{\kappa}(y,p_{2}) \bigg] \, ,
\end{eqnarray}
using the property $D_{\mu\nu}(x-y)=D_{\mu\nu}(y-x)$. Now plug in the Eqs. (\ref{eq:B-4}) and (\ref{eq:B-5}), express $T_\mu(x)$ in terms of its Fourier transform, and perform the $\int d{\bf x}$ and $\int dy$ integrals to get
\begin{eqnarray}
	\label{eq:B-9}
		\nonumber \phantom{}^{2}\hat H_{\text{int}}(t)&=&-\frac{i(2\pi)^7}{2}(2g)^2\int dk\, d{\bf p}_1\,d{\bf p}_2\, \frac{\epsilon^{\mu \nu \alpha \beta}\epsilon^{\rho \sigma \lambda \kappa }\tilde T_\nu( {l_1}) \tilde T_\sigma( {l_2})  }{k^2+i\epsilon}  (-ig_{\mu\rho})\big( p_{1\alpha}(p_{2}- k)_{\lambda} - k_\alpha (p_2 - k)_\lambda \big)\\
		&\times & \bigg[ e^{-i((p_1)^0-({l_1})^0+ k^0)t}\delta^3 ({\bf p}_1-{{\bf l}_1}+ {\bf k})\delta^4 (p_2+{l_2} + k) \epsilon^* _{s\kappa}(p_2) \epsilon _{r\beta}(p_1)\hat a_s^\dagger (p_2)\hat a_r(p_1)\\
\nonumber &&+   e^{i((p_1)^0+{l_1}^0+ k_0  )t} \delta ^3({\bf p}_1+{\bf l}_1+{\bf k})\delta ^4({ p}_2-{ l}_2+{ k}) \epsilon^* _{s\beta}(p_1) \epsilon _{r\kappa}(p_2)\hat a_s^\dagger (p_1)\hat a_r(p_2) \bigg] \,.
\end{eqnarray}
Next perform the $k$ integral:
\begin{eqnarray}
	\label{eq:B-10}
		 \phantom{}^{2}\hat H_{\text{int}}(t)&=&-\frac{i(2\pi)^3}{2}(2g)^2\int d{\bf p}_1 d{\bf p}_2\, d{ l}_1 d{ l}_2\, \epsilon^{\mu \nu \alpha \beta}\epsilon^{\rho \sigma \lambda \kappa }\tilde T_\nu( {l_1}) \tilde T_\sigma( {l_2}) (-ig_{\mu\rho}) \\
	\nonumber	&\times &  \bigg[\big( ( p_{1}+{p_2}+{l_2})_\alpha (2p_2 +{l_2}) _\lambda \big) e^{-i((p_1)^0-({l_1})^0-({p_2})^0-({l_2})^0)t}\delta^3 ({\bf p}_1-{{\bf l}_1}-{\bf p}_2-{{\bf l}_2})\frac{\epsilon^* _{s\kappa}(p_2) \epsilon _{s\beta}(p_1)\hat a_s^\dagger (p_2)\hat a_r(p_1)}{(p_2+l_2)^2+i\epsilon}\\
\nonumber &&+\big( (p_{1}+ p_2 -l_2)_\alpha (2p_2 -l_2)_\lambda  \big)   e^{i((p_1)^0+{l_1}^0- (p_2)^0+{l_2}^0  )t}  \delta ^3({\bf p}_1+{\bf l}_1-{\bf p}_2+{\bf l}_2)\frac{\epsilon^* _{s\beta}(p_1) \epsilon _{r\kappa}(p_2)\hat a_s^\dagger (p_1)\hat a_r(p_2)}{(p_2-l_2)^2+i\epsilon}\bigg] \,.
\end{eqnarray}

We are now in the position to compute the commutator of $\phantom{}^{2}\hat H_{\text{int}}(t)$ and 
$\mathcal{\hat D}_{uv}({\bf q})$ necessary for the refractive term of Eq.~(\ref{eq:3.27}). This is given by
\begin{eqnarray}
	\label{eq:B-11}
		  \left[ \phantom{}^{2}\hat H_{\text{int}}(t),\mathcal{\hat D}_{uv}({\bf q})\right] &=&-\frac{i(2\pi)^3}{2}(2g)^2\int d{\bf p}_1 d{\bf p}_2\, d{ l}_1 d{ l}_2\, \epsilon^{\mu \nu \alpha \beta}\epsilon^{\rho \sigma \lambda \kappa }\tilde T_\nu( {l_1}) \tilde T_\sigma( {l_2}) (-ig_{\mu\rho})e^{i({l_1}^0+{l_2}^0  )t} \\
	\nonumber	 && \qquad \times   \bigg[\big( ( p_{1}+{p_2}+{l_2})_\alpha (2p_2 +{l_2}) _\lambda \big) e^{-i((p_1)^0-({p_2})^0)t}\delta^3 ({\bf p}_1-{{\bf l}_1}-{\bf p}_2-{{\bf l}_2})\frac{\epsilon^* _{s\kappa}(p_2) \epsilon _{s\beta}(p_1)}{(p_2+l_2)^2+i\epsilon}\\
\nonumber &&\qquad\qquad \times(2\pi )^32q^0\left( \delta^{3}({\bf q} -  {\bf p}_1 )\delta_{ur}\hat a_s^\dagger ( p_2)\hat a_v (q)-\delta^{3}({\bf q} - {\bf p}_2 )\delta_{vs}\hat a_u^\dagger (q)\hat a_r ( p_1) \right)\\
\nonumber &&\qquad\qquad +\big( (p_{1}+ p_2 -l_2)_\alpha (2p_2 -l_2)_\lambda  \big)   e^{i((p_1)^0- (p_2)^0 )t}  \delta ^3({\bf p}_1+{\bf l}_1-{\bf p}_2+{\bf l}_2)\frac{\epsilon^* _{s\beta}(p_1) \epsilon _{r\kappa}(p_2)}{(p_2-l_2)^2+i\epsilon}\\
\nonumber &&\qquad\qquad\qquad\times(2\pi )^32q^0\left( \delta^{3}({\bf q} - {\bf p}_2 )\delta_{ur}\hat a_s^\dagger ( p_1)\hat a_v (q)-\delta^{3}({\bf q} - {\bf p}_1)\delta_{vs}\hat a_u^\dagger (q)\hat a_r ( p_2) \right)\bigg] \,.
\end{eqnarray}
Take the expectation value of Eq.~(\ref{eq:B-11}) above to get
\begin{eqnarray}
	\label{eq:B-12}
		  \left \langle \left[ \phantom{}^{2}\hat H_{\text{int}}(t),\mathcal{\hat D}_{uv}({\bf q})\right] \right \rangle &=&-\frac{i(2\pi)^3}{2}(2g)^2\int d{\bf p}_1 d{\bf p}_2\, d{ l}_1 d{ l}_2\, \epsilon^{\mu \nu \alpha \beta}\epsilon^{\rho \sigma \lambda \kappa }\tilde T_\nu( {l_1}) \tilde T_\sigma( {l_2}) (-ig_{\mu\rho})e^{i({l_1}^0+{l_2}^0  )t} \\
\nonumber &&\qquad \times(2\pi )^6(2q^0)^2\delta^{3}({\bf q} -  {\bf p}_1 )\delta^{3}({\bf p}_2 -  {\bf q} )\delta^3 ({\bf p}_1+{{\bf l}_1}-{\bf p}_2+{{\bf l}_2})\left( \delta_{ur}\rho_{sv}({\bf q})-\delta_{vs}\rho_{ur}({\bf q})  \right)  \\
		\nonumber  &&\qquad\qquad\times   \bigg[( p_{1}+{p_2}+{l_2})_\alpha (2p_2 +{l_2}) _\lambda  e^{-i((p_1)^0-({p_2})^0)t}\frac{\epsilon^* _{s\kappa}(p_2) \epsilon _{s\beta}(p_1)}{(p_2+l_2)^2+i\epsilon}\\
\nonumber &&\qquad\qquad\qquad + (p_{1}+ p_2 -l_2)_\alpha (2p_2 -l_2)_\lambda    e^{i((p_1)^0- (p_2)^0 )t} \frac{\epsilon^* _{s\beta}(p_1) \epsilon _{r\kappa}(p_2)}{(p_2-l_2)^2+i\epsilon} \bigg] \,.
\end{eqnarray}
Now perform the ${\bf p}_1$ and ${\bf p}_2$ integrals, giving 
\begin{eqnarray}
	\label{eq:B-13}
		 && \left \langle \left[ \phantom{}^{2}\hat H_{\text{int}}(t),\mathcal{\hat D}_{uv}({\bf q})\right] \right \rangle =-\frac{i(2\pi)^3}{2}(2g)^2\int d{ l}_1 d{ l}_2\, \delta^3 ({{\bf l}_1}+{{\bf l}_2})\epsilon^{\mu \nu \alpha \beta}\epsilon^{\rho \sigma \lambda \kappa }\tilde T_\nu( {l_1}) \tilde T_\sigma( {l_2}) (-ig_{\mu\rho})e^{i({l_1}^0+{l_2}^0  )t} \\
\nonumber &&\times \big( \delta_{ur}\rho_{sv}({\bf q})-\delta_{vs}\rho_{ur}({\bf q})  \big)  \bigg[( 2q+{l_2})_\alpha (2q +{l_2}) _\lambda  \frac{\epsilon^* _{s\kappa}(q) \epsilon _{s\beta}(q)}{(q+l_2)^2+i\epsilon} + (2q -l_2)_\alpha (2q -l_2)_\lambda    \frac{\epsilon^* _{s\beta}(q) \epsilon _{r\kappa}(q)}{(q-l_2)^2+i\epsilon} \bigg] \,.
\end{eqnarray}
Focus now on the following factors present in Eq.~(\ref{eq:B-13}):
\begin{eqnarray*}
&&\epsilon^{\mu \nu \alpha \beta}\epsilon_\mu\phantom{}^{ \sigma \lambda \kappa }\tilde T_\nu( {l_1}) \tilde T_\sigma( {l_2})  \epsilon^* _{s\kappa}(q) \epsilon _{r\beta}(q)( 2q\pm{l_2})_\alpha (2q \pm{l_2}) _\lambda =2\epsilon^{\mu \nu \alpha \beta}\epsilon_\mu\phantom{}^{ \sigma \lambda \kappa }\tilde T_\nu( {l_1}) \tilde T_\sigma( {l_2})  \epsilon^* _{s\kappa}(q) \epsilon _{r\beta}(q)( 2q\pm{l_2})_\alpha q _\lambda \,, 
\end{eqnarray*}
where we have used ${l}_{[\lambda}\tilde T_{\sigma]}( {l}) =0$ in arriving at the expression on the right. Now use the epsilon identity $\epsilon^{\mu \nu \alpha \beta}\epsilon_{\mu \sigma \lambda \kappa }= -6\delta^{[\nu}_\sigma \delta^\alpha _\lambda\delta^{\beta]} _\kappa$ to perform the contractions above and arrive at
\begin{eqnarray}
\label{eq:B-14}
&& -2\bigg([\tilde T_1\cdot \tilde T_2  ][(2q\pm l_2) \cdot q  ] [\epsilon_r \cdot \epsilon_s ] -[ \tilde T_2\cdot \tilde T_1 ][\epsilon_r \cdot q ] [(2q\pm l_2) \cdot \epsilon_s  ]+[\tilde T_2 \cdot \epsilon_r ][(2q\pm l_2) \cdot \epsilon_s  ] [\tilde T_1\cdot q ]\\
\nonumber &&\qquad- [\tilde T_2 \cdot \epsilon_r][(2q\pm l_2) \cdot q ] [\epsilon_s \cdot \tilde T_1 ] +[ (2q\pm l_2) \cdot \tilde T_2  ][q \cdot \epsilon_r  ] [\tilde T_1 \cdot \epsilon_s ]-[\tilde T_2\cdot (2q\pm l_2) ][\tilde T_1 \cdot q ] [\epsilon_r \cdot \epsilon_s  ]\bigg) \,.
\end{eqnarray}
To simplify this, use $\epsilon_r \cdot \epsilon_s = -\delta_{rs}$, which when contracted with $( \delta_{ur}\rho_{sv}({\bf q})-\delta_{vs}\rho_{ur}({\bf q}) )$ in Eq.~(\ref{eq:B-13}) vanishes, as well as $\epsilon_r(q) \cdot q = 0$ and $q\cdot q =0$. The remaining terms from the expression (\ref{eq:B-14}) then become
\begin{eqnarray*}
\label{eq:B-15}
\nonumber && \mp 2[\tilde T(l_2^0,{\bf l}_2) \cdot \epsilon_r ] \epsilon_{s\mu}q_\nu \bigg( (l_2)^\mu \tilde T^\nu(l_1^0,-{\bf l}_2) -(l_2)^\nu \tilde T^\mu (l_1^0,-{\bf l}_2)  \bigg)\\
&&\qquad=\pm 2[\tilde T(l_2^0,{\bf l}_2) \cdot \epsilon_r ]\epsilon_{s i} q_0 \bigg( (l_2)^i\tilde T^0(l_1^0,{\bf l}_2) +(l_2)^0\tilde T^i (l_1^0,{\bf l}_2)  \bigg)=\pm 2[\tilde T(l_2^0,{\bf l}_2) \cdot \epsilon_r ]\epsilon_{s i} \tilde T^i(l_1^0,{\bf l}_2) q_0\big( l_1^0 +l_2^0 \big) \, ,
\end{eqnarray*}
where we have used Eq.~(\ref{eq:4.36}) as well as repeated use of ${l}_{[\lambda}\tilde T_{\sigma]}( {l}) =0$.  Inserting this expression back into Eq.~(\ref{eq:B-13}), we have 
\begin{eqnarray}
	\label{eq:B-16}
		 && \left \langle \left[ \phantom{}^{2}\hat H_{\text{int}}(t),\mathcal{\hat D}_{uv}({\bf q})\right] \right \rangle =-(2\pi)^3(2g)^2\big( \delta_{ur}\rho_{sv}({\bf q})-\delta_{vs}\rho_{ur}({\bf q})  \big)  \int d{ l}_1 d{ l}_2\, \delta^3 ({{\bf l}_1}+{{\bf l}_2}) e^{i({l_1}^0+{l_2}^0  )t} \\
\nonumber &&\qquad\qquad\times [\epsilon_r \cdot \tilde T(l_2^0,{\bf l}_2)] [\tilde T(l_1^0,{\bf l}_2) \cdot \epsilon_s  ][ q\cdot (l_1 +l_2)]\bigg[\frac{1}{(l_2)^2+2l_2\cdot q+i\epsilon}- \frac{1}{(l_2)^2-2l_2\cdot q+i\epsilon}\bigg] \,.
\end{eqnarray}

We now define a quantity $\mathcal{T}_{ij}$ by
\begin{eqnarray}
	\label{eq:B-17}
		(2\pi)^3\delta^3(0) \mathcal{T}_{i j}(q)&=& -\int d{ l}_1 d{ l}_2\, (2\pi)^3 \delta^3 ({{\bf l}_1}+{{\bf l}_2})\tilde T_j(l_2^0,{\bf l}_2)\tilde T_i(l_1^0,{\bf l}_2)  [ q\cdot (l_1 +l_2)]\\
\nonumber &&\qquad\qquad \times\bigg[\frac{1}{(l_2)^2+2l_2\cdot q+i\epsilon}- \frac{1}{(l_2)^2-2l_2\cdot q+i\epsilon}\bigg].
\end{eqnarray}
Equation~(\ref{eq:B-16}) can be expressed conveniently in terms of this quantity: 
\begin{eqnarray}
	\label{eq:B-18}
		  i\left \langle \left[ \phantom{}^{2}\hat H_{\text{int}}(0),\mathcal{\hat D}_{uv}({\bf q})\right] \right \rangle &=&i(2\pi)^3\delta^3(0) (2g)^2\big( \delta_{ur}\rho_{sv}({\bf q})-\delta_{vs}\rho_{ur}({\bf q})  \big)  \epsilon_s^\mu(q) \epsilon_r^\nu(q)  \mathcal{T}_{\mu \nu}(q) \, ,
\end{eqnarray}
where we have exploited the fact that $\epsilon_s^i(q) \epsilon_r^j(q)  \mathcal{T}_{i j}(q)=\epsilon_s^\mu(q) \epsilon_r^\nu(q)  \mathcal{T}_{\mu \nu}(q)$ since $\epsilon(q)$ is purely spatial. Explicitly, the components of $i\left \langle \left[ \phantom{}^{2}\hat H_{\text{int}}(t),\mathcal{\hat D}_{uv}({\bf q})\right] \right \rangle$ are given by
\begin{eqnarray}
	\label{eq:B-19}
		  i\left \langle \left[ \phantom{}^{2}\hat H_{\text{int}}(t),\mathcal{\hat D}_{11}({\bf q})\right] \right \rangle &=&i(2\pi)^3\delta^3(0) (2g)^2\mathcal{T}_{\mu \nu}(q)\big( -\epsilon_1^\mu(q)\epsilon_2^\nu(q)\rho_{12}({\bf q})+ \epsilon_2^\mu(q)\epsilon_1^\nu(q)\rho_{21}({\bf q})  \big)\, ,    \\
	\label{eq:B-20}
		  i\left \langle \left[ \phantom{}^{2}\hat H_{\text{int}}(t),\mathcal{\hat D}_{12}({\bf q})\right] \right \rangle  &=&i(2\pi)^3\delta^3(0) (2g)^2    \mathcal{T}_{\mu \nu}(q)\big( [\epsilon_1^\mu(q)\epsilon_1^\nu(q)-\epsilon_2^\mu(q)\epsilon_2^\nu(q)]\rho_{12}({\bf q})\\
\nonumber &&\qquad\qquad\qquad\qquad\qquad\qquad -\epsilon_2^\mu(q)\epsilon_1^\nu(q)[\rho_{11}({\bf q})-\rho_{22}({\bf q})]  \big)\, ,\\
	\label{eq:B-21}
		  i\left \langle \left[ \phantom{}^{2}\hat H_{\text{int}}(t),\mathcal{\hat D}_{21}({\bf q})\right] \right \rangle &=&i(2\pi)^3\delta^3(0) (2g)^2    \mathcal{T}_{\mu \nu}(q)\big(  -[ \epsilon_1^\mu(q)\epsilon_1^\nu(q)-\epsilon_2^\nu(q)\epsilon_2^\mu(q)]\rho_{21}({\bf q})\\
\nonumber &&\qquad\qquad\qquad\qquad\qquad\qquad +\epsilon_1^\mu(q)\epsilon_2^\nu(q)[\rho_{11}({\bf q}) - \rho_{22}({\bf q})] \big) \, ,\\
	\label{eq:B-22}
		  i\left \langle \left[ \phantom{}^{2}\hat H_{\text{int}}(t),\mathcal{\hat D}_{22}({\bf q})\right] \right \rangle &=&i(2\pi)^3\delta^3(0) (2g)^2    \mathcal{T}_{\mu \nu}(q)\big( \epsilon_1^\mu(q)\epsilon_2^\nu(q) \rho_{12}({\bf q})-\epsilon_2^\mu(q)\epsilon_1^\nu(q)\rho_{21}({\bf q})  \big) \, .
\end{eqnarray}

\renewcommand{\theequation}{C-\arabic{equation}}
  \setcounter{equation}{0}  
\section{Gauge Invariance}
\label{sec-C}

Here we verify that the  calculation in Appendix B is gauge invariant by explicitly using a different gauge-fixed photon propagator, namely
\begin{equation}
\label{eq:B-23}
		D_{\mu\nu}(x-y)= \int\frac{d^4k}{(2\pi)^4}\frac{-ie^{-ik\cdot (x-y)}}{k^2+i\epsilon}\bigg( g_{\mu\nu}-(1-\xi )\frac{k_\mu k_\nu}{k^2} \bigg).
\end{equation}
Isolating the contribution to $\left \langle \left[ \phantom{}^{2}\hat H_{\text{int}}(t),\mathcal{\hat D}_{uv}({\bf q})\right] \right \rangle$ due to the term linear in $(1-\xi)$ gives
\begin{eqnarray}
	\label{eq:B-24}
		\nonumber && \left \langle \left[ \phantom{}^{2}\hat H_{\text{int}}(t)_\xi,\mathcal{\hat D}_{uv}({\bf q})\right] \right \rangle =-\frac{i(2\pi)^3}{2}(2g)^2\int d{ l}_1 d{ l}_2\, \delta^3 ({{\bf l}_1}+{{\bf l}_2})\epsilon^{\mu \nu \alpha \beta}\epsilon^{\rho \sigma \lambda \kappa }\tilde T_\nu( {l_1}) \tilde T_\sigma( {l_2}) (i(1-\xi ) )e^{i({l_1}^0+{l_2}^0  )t} \\
&&\times \big( \delta_{ur}\rho_{sv}({\bf q})-\delta_{vs}\rho_{ur}({\bf q})  \big)  \bigg[\frac{(q+l_2)_\mu (q+l_2)_\rho}{(q+l_2)^2}( q+q+{l_2})_\alpha (q+q +{l_2}) _\lambda  \frac{\epsilon^* _{s\kappa}(q) \epsilon _{s\beta}(q)}{(q+l_2)^2+i\epsilon} \\
\nonumber &&\qquad\qquad\qquad\qquad\qquad\qquad   + \frac{(q-l_2)_\mu (q-l_2)_\rho}{(q-l_2)^2}(q +q -l_2)_\alpha (q+q -l_2)_\lambda    \frac{\epsilon^* _{s\beta}(q) \epsilon _{r\kappa}(q)}{(q-l_2)^2+i\epsilon} \bigg] \,,
\end{eqnarray}
where we have performed all the integrals similar to those in arriving at Eq. (\ref{eq:B-13}). Simplify the above by making repeated use of the anti-symmetry of the epsilon tensor:
\begin{eqnarray*}
	\label{eq:B-25}
		\nonumber && \left \langle \left[ \phantom{}^{2}\hat H_{\text{int}}(t)_\xi,\mathcal{\hat D}_{uv}({\bf q})\right] \right \rangle =-\frac{i(2\pi)^3}{2}(2g)^2\int d{ l}_1 d{ l}_2\, \delta^3 ({{\bf l}_1}+{{\bf l}_2})\epsilon^{\mu \nu \alpha \beta}\epsilon^{\rho \sigma \lambda \kappa }\tilde T_\nu( {l_1}) \tilde T_\sigma( {l_2}) (i(1-\xi ) )e^{i({l_1}^0+{l_2}^0  )t} \\
&&\times \big( \delta_{ur}\rho_{sv}({\bf q})-\delta_{vs}\rho_{ur}({\bf q})  \big)  \bigg[\frac{(q+l_2)_\mu (q+l_2)_\rho}{(q+l_2)^2} q_\alpha q_\lambda  \frac{\epsilon^* _{s\kappa}(q) \epsilon _{s\beta}(q)}{(q+l_2)^2+i\epsilon} + \frac{(q-l_2)_\mu (q-l_2)_\rho}{(q-l_2)^2}q_\alpha q_\lambda    \frac{\epsilon^* _{s\beta}(q) \epsilon _{r\kappa}(q)}{(q-l_2)^2+i\epsilon} \bigg] , \\
		\nonumber && \qquad\qquad\qquad\qquad\qquad =-\frac{i(2\pi)^3}{2}(2g)^2\int d{ l}_1 d{ l}_2\, \delta^3 ({{\bf l}_1}+{{\bf l}_2})\epsilon^{\mu \nu \alpha \beta}\epsilon^{\rho \sigma \lambda \kappa }\tilde T_\nu( {l_1}) \tilde T_\sigma( {l_2}) (i(1-\xi ) )e^{i({l_1}^0+{l_2}^0  )t} \\
&&\qquad\qquad\qquad\qquad\times \big( \delta_{ur}\rho_{sv}({\bf q})-\delta_{vs}\rho_{ur}({\bf q})  \big)  \bigg[\frac{(l_2)_\mu (l_2)_\rho}{(q+l_2)^2} q_\alpha q_\lambda  \frac{\epsilon^* _{s\kappa}(q) \epsilon _{s\beta}(q)}{(q+l_2)^2+i\epsilon} + \frac{(l_2)_\mu (l_2)_\rho}{(q-l_2)^2}q_\alpha q_\lambda    \frac{\epsilon^* _{s\beta}(q) \epsilon _{r\kappa}(q)}{(q-l_2)^2+i\epsilon} \bigg] ,\\
		\nonumber && \qquad\qquad\qquad\qquad\qquad =-\frac{i(2\pi)^3}{2}(2g)^2\int d{ l}_1 d{ l}_2\, \delta^3 ({{\bf l}_1}+{{\bf l}_2}) (i(1-\xi ) )\big( \delta_{ur}\rho_{sv}({\bf q})-\delta_{vs}\rho_{ur}({\bf q})  \big)e^{i({l_1}^0+{l_2}^0  )t} \\
&&\qquad\qquad\qquad\qquad\times \bigg(\epsilon^{\mu \nu \alpha \beta}  (l_2)_\mu \tilde T_\nu( {l_1}) q_\alpha \bigg)\bigg( \epsilon^{\rho \sigma \lambda \kappa }(l_2)_\rho \tilde T_\sigma( {l_2}) q_\lambda \bigg)  \bigg[ \frac{\epsilon^* _{s\kappa}(q) \epsilon _{s\beta}(q)}{(q+l_2)^2(q+l_2)^2+i\epsilon} +   \frac{\epsilon^* _{s\beta}(q) \epsilon _{r\kappa}(q)}{(q-l_2)^2(q-l_2)^2+i\epsilon} \bigg] , \\
&&\qquad\qquad\qquad\qquad = 0.
\end{eqnarray*}
This expression vanishes because $ \epsilon^{\rho \sigma \lambda \kappa }(l_2)_\rho \tilde T_\sigma( {l_2} )=0 $, as we have required in order to arrive at Eq. (\ref{eq:B-16}). Therefore, the final evolution equations for the photon density matrix $\rho$ will indeed be independent of the gauge parameter $\xi$.

\bibliographystyle{apsrev}
\bibliography{references}

\end{document}